\newif\ifsinglecol
\singlecolfalse
\ifsinglecol
  \pdfoutput=1
  \documentclass[amsmath,amssymb,aps,prx,preprint,superscriptaddress]{revtex4-2}
  \newcommand{\colbreak}{}
\else
  \documentclass[amsmath,amssymb,aps,prx,reprint,superscriptaddress]{revtex4-2}
  \newcommand{\colbreak}{\\&\quad}
\fi
\ifsinglecol\newcommand{\colw}{86.8mm}\else\newcommand{\colw}{\linewidth}\fi
\newlength{\figlw}
\usepackage{capt-of}
\ifsinglecol
  \newcommand{\appfigbegin}{\begin{center}}\newcommand{\appfigend}{\end{center}\clearpage}
  \newcommand{\appcaption}{\captionof{figure}}
\else
  \newcommand{\appfigbegin}{\begin{figure*}}\newcommand{\appfigend}{\end{figure*}}
  \newcommand{\appcaption}{\caption}
\fi
\usepackage{amsmath}
\usepackage[table]{xcolor}
\makeatletter
\def\@hangfrom@section#1#2#3{\@hangfrom{#1#2}#3}%
\def\@hangfroms@section#1#2{#1#2}%
\makeatother
\usepackage{amsfonts,amssymb}
\usepackage{graphicx}
\makeatletter
\newif\iftwocol
\@ifclasswith{revtex4-2}{reprint}{\twocoltrue}{\twocolfalse}
\makeatother
\usepackage{hyperref}
\usepackage{bookmark}
\usepackage{cleveref}
\AtBeginDocument{\let\eprint\relax
  \newcommand{\eprint}[2][]{Preprint at \url{https://arxiv.org/abs/#2}}}
\crefname{figure}{Fig.}{Figs.}
\Crefname{figure}{Figure}{Figures}
\crefname{equation}{Eq.}{Eqs.}
\Crefname{equation}{Equation}{Equations}
\crefname{section}{Sec.}{Secs.}
\Crefname{section}{Section}{Sections}
\crefname{table}{Table}{Tables}
\usepackage{academicons}
\usepackage{multirow}
\usepackage[version=4]{mhchem}
\usepackage{makecell}
\usepackage{pgf}
\usepackage{booktabs}
\usepackage[english]{babel}
\usepackage{tikz}
\ifsinglecol
  \usepackage[mathlines]{lineno}
  \linenumbers
  \makeatletter
  \renewcommand\section{\@startsection{section}{1}{\z@}%
    {0.8cm \@plus1ex \@minus .2ex}{0.5cm}{\normalfont\large\bfseries}}
  \renewcommand\subsection{\@startsection{subsection}{2}{\z@}%
    {.8cm \@plus1ex \@minus .2ex}{.5cm}{\normalfont\normalsize\bfseries}}
  \renewcommand\subsubsection{\@startsection{subsubsection}{3}{\z@}%
    {.8cm \@plus1ex \@minus .2ex}{.5cm}{\normalfont\normalsize\itshape}}
  \makeatother
\fi

\newcommand{\panelw}[3]{%
  \begin{tikzpicture}
    \node[anchor=south west,inner sep=0] (im) {\includegraphics[width=#1]{#2}};
    \node[anchor=north west,font=\bfseries,inner sep=1pt] at (im.north west) {#3};
  \end{tikzpicture}}

\newcommand{\CellMin}{0}
\newcommand{\CellMax}{0.4}
\colorlet{cmin}{green!10!white}
\colorlet{cmed}{yellow!20!white}
\colorlet{cmax}{red!20!white}
\newcommand{\ColorCell}[1]{%
  \begingroup
  \pgfmathsetmacro{\val}{#1}%
  \pgfmathsetmacro{\norm}{(\val-\CellMin)/(\CellMax-\CellMin)}%
  \pgfmathsetmacro{\norm}{max(min(\norm,1),0)}%
  \pgfmathsetmacro{\pct}{\norm*100}%
  \pgfmathtruncatemacro{\pctint}{round(\pct)}%

  \ifnum\pctint<50
    \pgfmathtruncatemacro{\mix}{round(100 - 2*\pctint)}%
    \edef\temp{\noexpand\cellcolor{cmin!\mix!cmed}}
  \else
    \pgfmathtruncatemacro{\mix}{round(200 - 2*\pctint)}%
    \edef\temp{\noexpand\cellcolor{cmed!\mix!cmax}}%
  \fi
  \temp #1%
  \endgroup
}

\makeatletter
\def\@fnsymbol#1{%
  \ifcase#1\or \ensuremath{\dagger}\or \ensuremath{\ddagger}\or
  \mathsection\or \mathparagraph\or \|\or **\or
  \ensuremath{\dagger\dagger}\or \ensuremath{\ddagger\ddagger}\else\@ctrerr\fi}
\makeatother

\begin{document}

\title{Learning \textit{ab initio} phase-field models}

\author{Mengyi Chen}
\affiliation{Department of Mathematics, National University of Singapore, Singapore}

\author{Peichen Zhong}
\affiliation{Department of Materials Science and Engineering, National University of Singapore, Singapore}
\affiliation{Institute for Functional Intelligent Materials, National University of Singapore, Singapore}

\author{Zihan Zhang}
\affiliation{Department of Materials Science and Engineering, National University of Singapore, Singapore}

\author{Qianxiao Li}
\email{qianxiao@nus.edu.sg}
\affiliation{Department of Mathematics, National University of Singapore, Singapore}
\affiliation{Institute for Functional Intelligent Materials, National University of Singapore, Singapore}

\date{\today}

\date{\today}

\begin{abstract}

Simulating microstructure evolution requires quantum-mechanical accuracy and mesoscopic reach in length and time scales, a combination that no current method achieves. Classical phase-field models provide this reach, but their accuracy is limited by phenomenological free energies and mobilities. Here we develop a framework for learning \textit{ab initio} phase-field models, where the mesoscopic equation is not postulated but derived from a Mori--Zwanzig projection of molecular dynamics onto species-density fields under explicit assumptions. The nonlocal free energy and mobility left unspecified by this equation are parametrized by neural networks and learned from short molecular dynamics trajectories generated with machine-learning interatomic potentials of \textit{ab initio} accuracy. We demonstrate the framework on an iron--boron melt and on hydrogen--helium mixtures under planetary conditions. For iron--boron, the model shows that the melt at the FeB$_4$ composition is spinodally unstable at ambient pressure but stabilized at 10\,GPa, offering a thermodynamic rationale for why FeB$_4$ has been synthesized only under high pressure. For hydrogen--helium, the model predicts the immiscibility boundary and captures droplet nucleation and growth in helium-rain simulations of a column corresponding to 2.2 million atoms, far beyond the scale of atomistic modeling at comparable accuracy. Trained across compositions and conditions, such models could provide a mesoscopic counterpart to \textit{ab initio} molecular dynamics.

\end{abstract}


\maketitle
\ifsinglecol\clearpage\fi

\begin{figure*}[!t]
    \centering
    \includegraphics[width=1.0\linewidth]{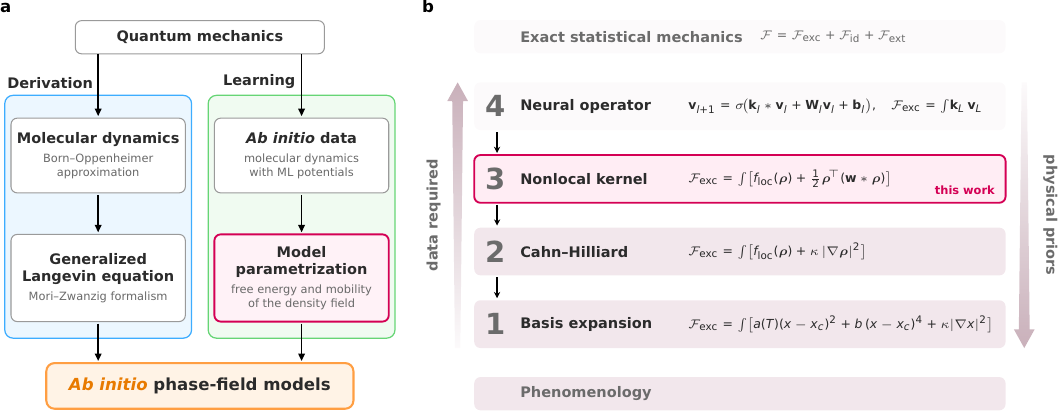}
    \caption{\textbf{\textit{Ab initio} phase-field models.}
    \textbf{a}, The proposed \textit{ab initio} phase-field (AIPF) model is \textit{ab initio} in both its derivation and its learning, which start from quantum mechanics. The derivation proceeds from quantum mechanics to molecular dynamics under the Born--Oppenheimer approximation, then to a generalized Langevin equation through the Mori--Zwanzig formalism, and finally to the \textit{ab initio} phase-field model under the stated assumptions. For the learning process, the AIPF
    model is trained on molecular dynamics trajectories generated with a machine-learning interatomic potential (MLIP) of \textit{ab initio} accuracy. The free energy and mobility of the density field are parametrized by neural networks, with the free-energy functional chosen from the ladder in \textbf{b}.
    \textbf{b}, Ladder of free-energy functionals from exact statistical
    mechanics to phenomenology, where $\mathcal F$ splits into the known
    ideal-gas and external-potential parts and the unknown excess part
    $\mathcal F_{\mathrm{exc}}$. Each lower rung introduces additional restrictions, reducing flexibility and training-data requirements. 
    Rung 4 is a neural operator, with layer fields $\mathbf v_l$ starting from
    $\mathbf v_0=\boldsymbol\rho$, convolution kernels
    $\mathbf k_l(\mathbf r-\mathbf r',T)$, channel-mixing
    weights $\mathbf W_l$, biases $\mathbf b_l(T)$ and a pointwise
    nonlinearity $\sigma$. Rung 3, adopted in this work, restricts
    the excess free energy to a local free-energy density
    $f_{\mathrm{loc}}(\boldsymbol\rho,T)$ plus a nonlocal kernel
    $\mathbf W(\lvert\mathbf r-\mathbf r'\rvert,T)$. Rung 2 approximates the nonlocal term by the Cahn--Hilliard square-gradient form with coefficient $\kappa$. Rung 1 further restricts the local free energy to a prescribed basis expansion in composition $x$, such as Landau polynomials or Redlich--Kister expansions, with coefficients $c_n(T)$.}
    \label{fig:ladders}
\end{figure*}

Collective behavior in materials, including nucleation, phase separation and coarsening, emerges from microscopic interactions and spans length scales from nanometers to micrometers and time scales from nanoseconds to hours~\cite{liCurrentinducedTransitionParticlebyparticle2014,maoLightinducedReversalIon2021}.
Accurate and scalable simulation across these length and time scales is therefore central to understanding the underlying physical mechanisms.
\textit{Ab initio} atomistic simulations retain quantum-mechanical accuracy, but their computational cost limits access to these length and time scales~\cite{jiaPushingLimitMolecular2020}. 
Phase-field models enable scalable simulations by representing microstructure through continuous fields at mesoscopic scales and have been successful across a wide range of materials processes~\cite{hohenberg1977theory,karmaQuantitativePhasefieldModeling1998,chenPhaseFieldModelsMicrostructure2002,steinbachPhasefieldModelsMaterials2009a,tourretPhasefieldModelingMicrostructure2022}.
However, they rely on prescribed forms of the free energy and mobility, with a few coefficients fitted to experiments or to the very phenomena being predicted, resulting in a phenomenological description with limited predictive accuracy~\cite{bhattacharyyaFirstprinciplesPhaseField2019a,tourretPhasefieldModelingMicrostructure2022,pruessnerPhasefieldModelCrystal2008}.

Accurate and scalable mesoscopic simulation therefore remains a grand challenge, motivating the development of \textit{ab initio} phase-field (AIPF) models grounded in microscopic theory and data (\cref{fig:ladders}).
We use the term \textit{ab initio} in the same spirit as in \textit{ab initio} molecular dynamics (AIMD).
In AIMD, the equations of nuclear motion follow from quantum mechanics under the Born--Oppenheimer approximation.
The forces are computed using density functional theory (DFT), where the exchange--correlation functional is chosen from a ladder of systematically improvable approximations known as Jacob's ladder~\cite{perdewJacobsLadderDensity2001}.
Similarly, for phase-field models to qualify as \textit{ab initio}, we impose two corresponding requirements.
First, the evolution equation is derived from microscopic dynamics under stated assumptions, without phenomenological prescriptions (\cref{fig:ladders}a).
Second, the free-energy functional and mobility are selected from ladders of systematically improvable models, with their parameters learned from microscopic data of \textit{ab initio}
accuracy (free-energy ladder in \cref{fig:ladders}b).

Several attempts have been made to obtain the mesoscopic free energy from microscopic theory rather than from the phenomena it describes, and each
occupies a rung of the free-energy ladder.
At the lowest rung, phase-field models retain a prescribed basis expansion for the free energy and determine its few coefficients from static
DFT calculations~\cite{bhattacharyyaFirstprinciplesPhaseField2019a,vaithyanathanMultiscaleModelingPrecipitate2002}.
At the next rung of the ladder, the basis expansion for the local free energy is replaced by a neural network trained on equilibrium chemical potentials~\cite{teichertMachineLearningMaterials2019a,shojaeiBridgingScalesMachine2024}.
At both rungs, only the local free energy is obtained from microscopic calculations.
The gradient coefficient that sets the interfacial energy and the mobility that sets the relaxation rate are fixed separately or prescribed.
Neural classical density functional theory (cDFT) extends this hierarchy to nonlocal free-energy functionals learned from equilibrium data~\cite{sammullerNeuralFunctionalTheory2023,buiUnifiedMachineLearning2026,chengEquivariantLearningTransferable2026}.
Learning these functionals typically requires hundreds to thousands of simulations under external potentials to sample diverse equilibrium
profiles~\cite{sammullerNeuralFunctionalTheory2023,buiUnifiedMachineLearning2026,chengEquivariantLearningTransferable2026}.
The computational cost restricts applications largely to simple model fluids, such as hard spheres and Lennard-Jones fluids, in planar geometries.
Few extend to three-dimensional density fields~\cite{chengEquivariantLearningTransferable2026}, or to complex liquids such as water, using training data generated with machine-learning interatomic potentials (MLIPs)~\cite{buiUnifiedMachineLearning2026}.
The learned equilibrium free-energy functionals have also been used to predict dynamics within dynamical density functional theory, under the adiabatic approximation in which the equilibrium functional determines the instantaneous driving forces~\cite{ramLearnedFreeenergyFunctionals2025}.
A few studies instead learn the free energy and mobility jointly from dynamics, but use lattice or chain models rather than \textit{ab initio}
data~\cite{huangStatisticalPhysicsInformedNeural2025,liHypothesisdrivenConstructionMesoscopic2026}.
All of these approaches seek an explicit mesoscopic equation. Multiscale coupling instead bypasses the mesoscopic equation and calls the microscopic solver at every macroscopic step~\cite{eHeterogeneousMultiscaleMethods2003,kevrekidisEquationfreeCoarsegrainedMultiscale2003}, so every new simulation still requires microscopic runs.

\begin{figure*}[!t]
    \centering
    \includegraphics[width=1.0\linewidth]{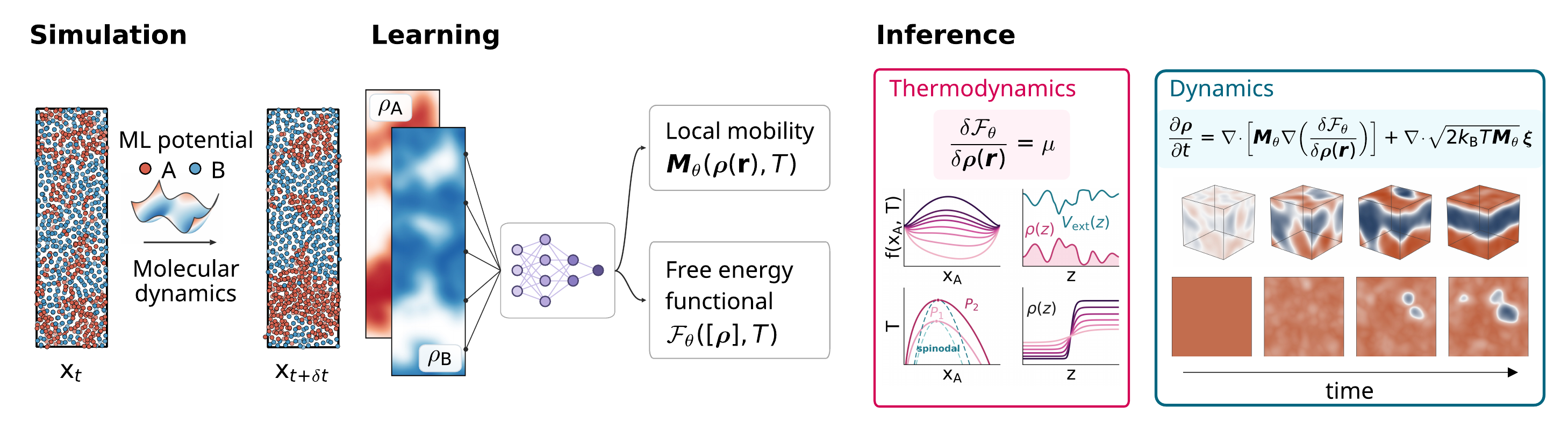}
    \caption{\textbf{Framework for learning \textit{ab initio} phase-field models.}
    In the simulation stage, molecular dynamics with an
    MLIP propagates a configuration $\mathrm x_t$ over
    a short interval $\delta t$.
    In the learning stage, each configuration is coarse-grained
    into species density fields $\boldsymbol\rho$,
    from which neural networks with trainable parameters $\theta$ output the local mobility $\mathbf M_\theta$ and
    the free-energy functional $\mathcal F_\theta$.
    At inference, the thermodynamics follows from
    minimizing the learned functional and the dynamics from integrating the
    mesoscopic equation.}
    \label{fig:framework}
\end{figure*}

Despite these advances, an \textit{ab initio} phase-field framework for accurately predicting both thermodynamics and dynamics remains to be established. 
Here, we present a framework for learning \textit{ab initio} phase-field models. The mesoscopic equation follows from molecular dynamics (MD) under three
assumptions: Markovian dynamics, local mobility and Gaussian fluctuating forces. The nonlocal free-energy functional and mobility are then learned jointly from MD trajectories generated
using MLIPs of \textit{ab initio} accuracy (\cref{fig:framework}).  Recent advances in MLIPs, including foundation potentials trained on extensive datasets, make the required training data computationally accessible~\cite{batatiaFoundationModelAtomistic2025,dengCHGNetPretrainedUniversal2023c}.
Training requires only short MD trajectories of systems containing a few thousand atoms, generated without external potentials, making the approach data efficient. The learned AIPF model can then simulate much larger systems over longer times.

\section*{\texorpdfstring{\textit{Ab initio}}{Ab initio} phase-field models}
\label{sec:model}

We consider a multicomponent system in which each species $\alpha$ comprises $N_\alpha$ atoms with positions $\mathbf r_j^\alpha(t)$ and momenta $\mathbf p_j^\alpha(t)$.
The dynamics is simulated using an MLIP of \textit{ab initio} accuracy.
We describe each species $\alpha$ by a band-limited number-density field $\rho_\alpha(\mathbf r,t)$.
This density field is constructed by removing modes with $|\mathbf k|>k_{\mathrm{cut}}$ from the Fourier representation of the microscopic particle density and transforming the remaining modes back to real space.
The cutoff sets the spatial resolution, retaining long-wavelength density variations associated with slow mesoscopic dynamics.
We apply an exact Mori--Zwanzig projection onto these band-limited density fields, yielding a generalized Langevin equation (GLE)~\cite{moriTransportCollectiveMotion1965,zwanzigMemoryEffectsIrreversible1961,grabert1982projection}.
The GLE describes the evolution of the density fields through a memory-dependent response to the free-energy gradient, while a fluctuating force accounts for stochastic effects from unresolved degrees of freedom.

Although exact, the GLE is difficult to evaluate because its memory kernel and fluctuating force are not directly accessible.
We make three assumptions to reduce the GLE to a practical mesoscopic model.
First, in many systems, such as simple fluids, fluid mixtures, and colloidal suspensions, a particle's velocity decorrelates within a few collisions, so the memory kernel decays far faster than the evolution of the resolved density modes~\cite{ortizdezarateHydrodynamicFluctuations2006,vrugtClassicalDynamicalDensity2020}.
We therefore take the dynamics to be Markovian, replacing the memory kernel by its time integral, which we call the mobility~\cite{zwanzigMemoryEffectsIrreversible1961,grabert1982projection,hijonMoriZwanzigFormalism2010a}.
Second, we restrict the mobility to depend only on the local densities, since fully nonlocal mobility is difficult to identify from data.
Third, we assume the fluctuating force to be Gaussian conditional on the current density fields, with covariance fixed by the fluctuation--dissipation relation~\cite{onsagerFluctuationsIrreversibleProcesses1953,ortizdezarateHydrodynamicFluctuations2006,hijonMoriZwanzigFormalism2010a}.
Under these three assumptions, the GLE reduces to the mesoscopic model (Methods, `Theoretical derivation'),
\begin{equation}
  \partial_t\boldsymbol\rho
  =\nabla\!\cdot\!\Big[\mathbf M\,\nabla\frac{\delta\mathcal F}{\delta\boldsymbol\rho}\Big]
  +\nabla\!\cdot\!\Big[\sqrt{2k_{\mathrm B}T\mathbf M}\,
    \boldsymbol\xi(\mathbf r,t)\Big],
  \label{eq:meso}
\end{equation}
with $\mathbf M=\mathbf M(\boldsymbol\rho(\mathbf r),T)$ the local mobility, $\mathcal F=\mathcal F([\boldsymbol\rho],T)$ the free-energy functional, and $\boldsymbol\xi$ a unit Gaussian white noise.
The dynamics admits the Boltzmann equilibrium distribution associated with $\mathcal{F}$, 
and the most probable configurations satisfy the stationarity condition
\begin{equation}
  \frac{\delta\mathcal F([\boldsymbol\rho],T)}{\delta\boldsymbol\rho}
    =\boldsymbol\mu,
  \label{eq:el}
\end{equation}
with $\boldsymbol\mu$ the constant chemical potentials.

\Cref{eq:meso} fixes the form of the equation but leaves $\mathcal F$ and $\mathbf M$ unspecified.
The free energy can be decomposed as $\mathcal F=\mathcal F_{\mathrm{exc}}+\mathcal F_{\mathrm{id}}+\mathcal F_{\mathrm{ext}}$, where the ideal-gas part $\mathcal F_{\mathrm{id}}$ and the external-potential part $\mathcal F_{\mathrm{ext}}$ are known exactly and the excess part $\mathcal F_{\mathrm{exc}}$ accounts for interparticle interactions.
We adopt rung 3 of \cref{fig:ladders}b for $\mathcal F_{\mathrm{exc}}$, 
\begin{equation}
\begin{aligned}[b]
  &\mathcal F_{\mathrm{exc}}([\boldsymbol\rho],T)
  =\int f_{\mathrm{loc}}(\boldsymbol\rho,T)\,\mathrm d\mathbf r \colbreak
  {}+\frac12\sum_{\alpha\beta}\iint\rho_\alpha(\mathbf r)\,W_{\alpha\beta}(\lvert\mathbf r-\mathbf r'\rvert,T)\,\rho_\beta(\mathbf r')\,\mathrm d\mathbf r\,\mathrm d\mathbf r',
\end{aligned}
  \label{eq:pair_b}
\end{equation}
with $f_{\mathrm{loc}}$ a function of the local densities and $W_{\alpha\beta}$ the radial kernel between species $\alpha$ and $\beta$.
Rung 3 balances nonlocal expressiveness with data efficiency.
Its nonlocal kernel extends the square-gradient term of rung 2 to the full wavevector dependence of the pair correlations, while its quadratic pair term provides a stronger physical prior than rung 4, reducing the amount of training data required.
Under the assumption of local mobility, $\mathbf M$ depends only on the densities at the same point.
As a parametrization choice, $\mathbf M$ is taken to be separable, a function of the densities scaled by a temperature factor.
The temperature factor takes an Arrhenius form with learned activation energies when necessary (Methods, `Model parametrization').

\begin{figure}[!t]
    \centering
    \panelw{\colw}{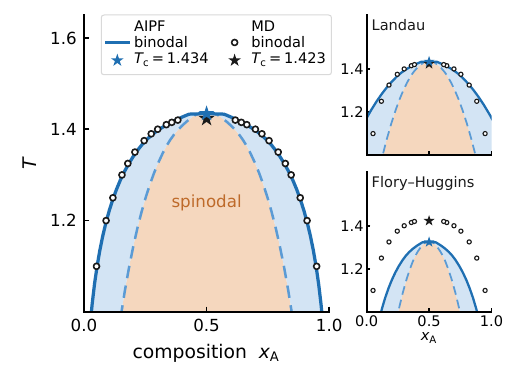}{a}\\[1.0ex]
    \panelw{\colw}{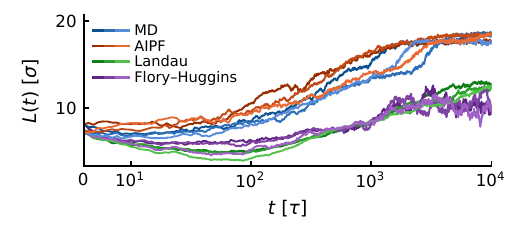}{b}
    \caption{\textbf{Binary Lennard-Jones mixture.}
\textbf{a}, Phase diagrams from the AIPF model and the Landau (upper small panel) and Flory--Huggins (lower small panel) baselines, compared with MD~\cite{dasStaticDynamicCritical2006}.
All models are trained on the same data using the same loss function and mobility parametrization, differing only in the form of the free-energy functional.
\textbf{b}, Domain size $L(t)$ at $T=1.20$ (3 seeds each), compared between MD and the three models, 
with $L(t)$ defined in Supplemental Material, Section S3\,A.
Fits to $L\propto t^\beta$ give effective exponents $\beta_{\mathrm{MD}}=0.225\pm0.022$ and $\beta_{\mathrm{AIPF}}=0.217\pm0.008$.
\label{fig:lj}}
\end{figure}

\begin{figure*}[!t]
    \centering
    \includegraphics[width=\linewidth]{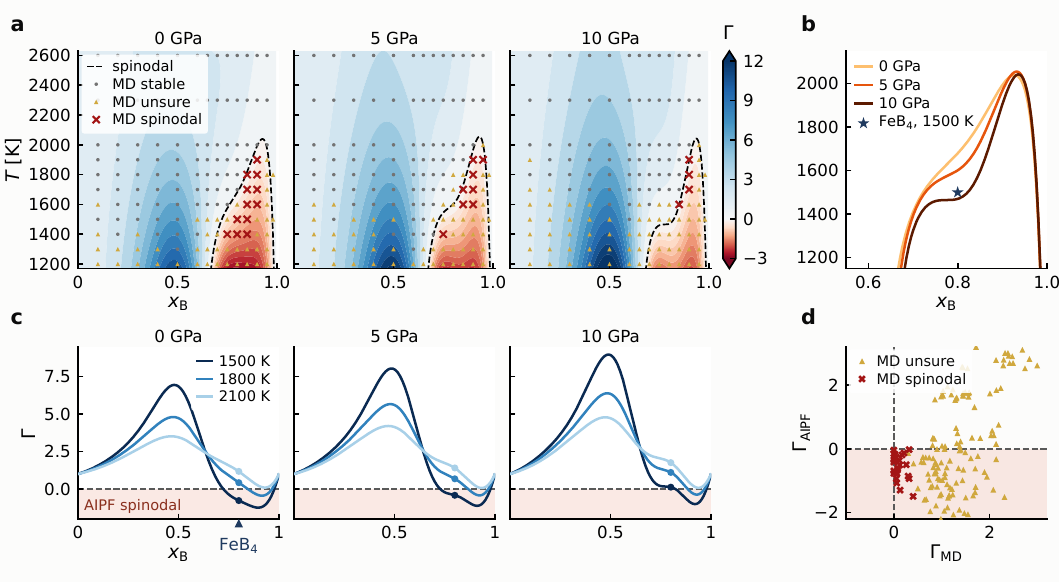}
    \caption{\textbf{Iron--boron melt.}
    \textbf{a}, The predicted thermodynamic factor $\Gamma(x_{\mathrm B},T)$ at 0, 5 and $10\,$GPa is shown with the MD classification of the sampled states. For each state, $\Gamma_{\mathrm{MD}}$ is obtained from a $50$--$500\,$ps trajectory by extrapolating the concentration structure factor $S_{cc}(k)$ to $k=0$ using the Ornstein--Zernike form. States are classified as stable when $\Gamma_{\mathrm{MD}}\ge1$, $S_{cc}(0)$ is converged to within $10\,\%$ and composition fluctuations show no growth, and as spinodal when $\Gamma_{\mathrm{MD}}\le0.35$ and the melt coarsens. All remaining states are classified as unresolved.
    \textbf{b}, The spinodal regions predicted by the AIPF model at the three pressures. The star marks the FeB$_4$ composition at $1500\,$K, inside the spinodal region at 0 and $5\,$GPa and outside it at $10\,$GPa.
    \textbf{c}, The thermodynamic factor $\Gamma(x_{\mathrm B})$ is shown at three temperatures for each pressure. Dots mark the FeB$_4$ composition.
    \textbf{d}, The predictions $\Gamma_{\mathrm{AIPF}}$ are compared with $\Gamma_{\mathrm{MD}}$ for states classified by MD as spinodal or unresolved. The AIPF model agrees with the spinodal classification from MD and predicts the sign of $\Gamma$ for unresolved states.
    \label{fig:feb}}
\end{figure*}

\begin{figure*}[!t]
    \centering
    \setlength{\figlw}{\ifsinglecol0.90\linewidth\else\linewidth\fi}%
    \includegraphics[width=\figlw]{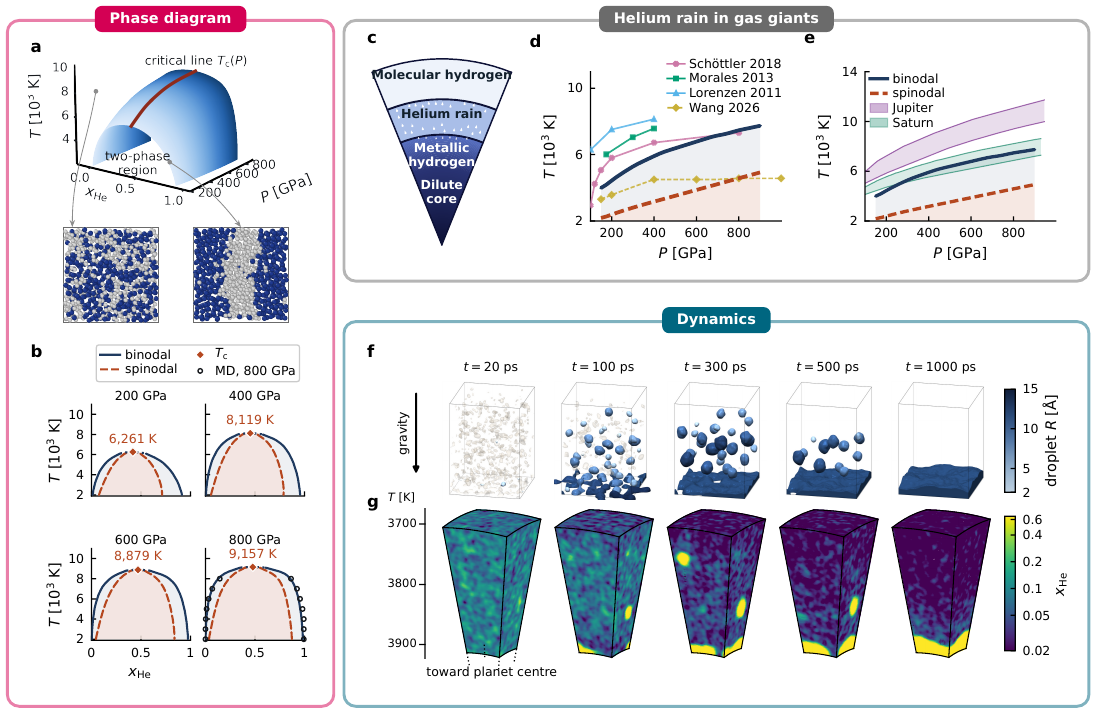}
    \caption{\textbf{Hydrogen--helium mixture.}
    The AIPF model is trained at 200, 400, 600 and $800\,$GPa, with predictions in \textbf{d,e} extending to $150$--$900\,$GPa.
    \textbf{a}, Binodal surface in $(x_{\mathrm{He}},P,T)$ with the
    critical line $T_{\mathrm c}(P)$.
    \textbf{b}, Isobaric sections of \textbf{a} at 200, 400, 600 and $800\,$GPa, with the predicted critical temperatures. Circles denote binodal compositions measured from MD at $800\,$GPa.
    \textbf{c}, Schematic of helium demixing from metallic hydrogen and settling toward the center of a gas giant. 
    \textbf{d}, Predicted binodal (solid) and spinodal (dashed) at the protosolar composition $x_{\mathrm{He}}=0.089$, compared with published binodals from Sch{\"o}ttler and Redmer~\cite{schottlerAbInitioCalculation2018}, Morales \textit{et al.}~\cite{moralesHydrogenheliumDemixingFirst2013} and Lorenzen \textit{et al.}~\cite{lorenzenMetallizationHydrogenheliumMixtures2011}, and the spinodal from Wang \textit{et al.}~\cite{wangHydrogenheliumImmiscibilityBoundary2026a}, computed with the vdW-DF functional.
    \textbf{e}, The same boundaries are compared with the interior profiles of Jupiter and Saturn. Bands span published isentropes and present-day thermal profiles compiled by Wang \textit{et al.}~\cite{wangHydrogenheliumImmiscibilityBoundary2026a}. Helium rain is expected wherever a profile lies below the binodal.
   \textbf{f,g}, Helium rain simulated for $1\,$ns in a $13.4\times13.4\times17.9\,$nm column corresponding to 2.2 million atoms, at $x_{\mathrm{He}}=0.089$ and $400\,$GPa (Supplemental Material, Section S3). Temperature increases from $3700\,$K at the top to $3900\,$K at the base, with enhanced gravity directed toward the planet's center.
    \textbf{f}, Opaque surfaces show helium droplets colored by radius. White surfaces indicate helium-rich fluctuations below the droplet threshold.
    \textbf{g}, The helium fraction is shown on a wedge section of the same domain at the same times.
    \label{fig:hhe_main}%
    }
\end{figure*}

With the model parametrization specified, we learn $\mathcal F_{\mathrm{exc},\theta}$ and $\mathbf M_\theta$ from short MD trajectories of systems containing only a few thousand atoms, generated without external potentials (\cref{fig:framework}; Methods, `Training data').
We consider two kinds of initial configuration, homogeneous mixtures and slabs with interfaces between regions of different composition.
We learn the dynamics from phase separation in the former and interface relaxation in the latter, where phase separation provides direct access to unstable states in the spinodal region.
We estimate the density evolution over an interval $\delta t$ longer than the microscopic memory time, so that a Markovian description applies, but short enough that the drift does not change appreciably.
We match the deterministic part of \cref{eq:meso} to the measured density evolution (Methods, `Training objective').
Fluctuations in homogeneous equilibrium trajectories provide additional constraints on the mobility and the curvature of the free energy (Methods, `Fluctuation-based measurements').

\section*{Results}
\noindent\textbf{Binary Lennard-Jones mixture.}  
As a preliminary check, we test the framework on a symmetric binary
Lennard-Jones mixture of species $A$ and $B$~\cite{dasStaticDynamicCritical2006}.
The two species interact through a pair potential with short-range repulsion and longer-range attraction, and evolve under overdamped Langevin dynamics.
Below the critical temperature, the mixture demixes spontaneously within the spinodal region, forming two phases with equilibrium compositions given by the binodal.
Between the binodal and spinodal boundaries, the mixture is metastable and phase separation requires nucleation.
The phase diagram follows from the bulk free energy $f(x_A,T)$, obtained by evaluating the learned free-energy functional on homogeneous states, where $x_A$ is the fraction of species $A$.
A common tangent to $f$ gives the binodal compositions, while $\partial^2 f/\partial x_A^2<0$ identifies the spinodal region.
For both the phase diagram (\cref{fig:lj}a) and the characteristic domain size $L(t)$ during spinodal decomposition (\cref{fig:lj}b), the AIPF model agrees closely with MD and outperforms the Landau and Flory--Huggins baselines (Methods, \cref{eq:landau,eq:fh}).
We further compare the AIPF model with MD and the two baselines for equilibrium composition profiles under external potentials (\cref{fig:appendix_lj_vext}) and interface relaxation in slabs (\cref{fig:appendix_lj_slab}).

\medskip
\noindent\textbf{Iron--boron melt.}
Computational searches identify promising solid-state materials by their stability on the formation-energy convex hull~\cite{merchantScalingDeepLearning2023,riebesellFrameworkEvaluateMachine2025}, yet thermodynamic stability alone does not guarantee synthesis, which also depends on the synthesis route and conditions.
For example, FeB$_4$ is predicted to be thermodynamically stable~\cite{kolmogorovNewSuperconductingSemiconducting2010}, but its synthesis has been reported only under high pressure~\cite{gouDiscoverySuperhardIron2013}.
One route to solid-state synthesis proceeds through a transient melt, formed at the interface between two solid precursors, from which the target crystal nucleates~\cite{zhangInterfacialmeltStabilityThermodynamic2026}.
Zhang \textit{et al.}\ proposed that synthesis through this route requires the melt at the target composition to be locally stable and, using MLIP-driven MD simulations, showed that the Fe--B melt near FeB$_4$ is spinodally unstable at ambient pressure but stabilized by high pressure~\cite{zhangInterfacialmeltStabilityThermodynamic2026}.

Identifying the spinodal region requires the composition curvature of the Gibbs free energy of mixing, $\Delta G_{\mathrm{mix}}$, of the homogeneous melt.
Inside this region, however, the homogeneous melt is unstable and out of equilibrium, whereas conventional methods rely on equilibrium
measurements of the homogeneous melt.
Prior works either compute the free energy from the enthalpy and entropy of a homogeneous melt in small cells that suppress phase separation~\cite{lorenzenMetallizationHydrogenheliumMixtures2011,moralesHydrogenheliumDemixingFirst2013,schottlerAbInitioCalculation2018,wahl2015high,zhangInterfacialmeltStabilityThermodynamic2026}, or obtain its curvature from equilibrium structure factors~\cite{chengComputingChemicalPotentials2022,wangHydrogenheliumImmiscibilityBoundary2026a,zhangInterfacialmeltStabilityThermodynamic2026}. 
The former requires the nonideal mixing entropy, which is hard to determine accurately. 
The latter yields a nonnegative curvature estimate by construction and identifies the spinodal only as this estimate approaches zero, with resolution limited by system size and simulation time.
Both yield results only at discrete state points, so obtaining a continuous phase boundary requires fitting a prescribed free-energy form, such as a Redlich--Kister expansion~\cite{schottlerAbInitioCalculation2018,wangHydrogenheliumImmiscibilityBoundary2026a, huangInitioFreeEnergies2022}.

The negative curvature, however, governs the early stage of phase separation. 
The demixing that
prior methods
either suppress or cannot use is itself informative about the curvature.
By learning from the dynamics, the AIPF model accesses the spinodal region directly.
Using the MACE potential of Zhang \textit{et al.}~\cite{zhangInterfacialmeltStabilityThermodynamic2026}, we learn a single AIPF model of the Fe--B melt at 0, 5 and 10\,GPa, where pressure enters implicitly through the model's dependence on species densities and temperature, as expressed by the Euler relation (Methods, \cref{eq:euler}).
The Gibbs free energy of mixing $\Delta G_{\mathrm{mix}}$ then follows by adding the pressure--volume term $PV$ to $\mathcal F_\theta$ (Methods, \cref{eq:dgmix}, and \cref{fig:appendix_feb_dgmix}).
Its curvature is quantified by the thermodynamic factor
\begin{equation}
  \Gamma = \frac{x_{\mathrm{B}}(1-x_{\mathrm{B}})}{k_{\mathrm{B}}T}\left(\frac{\partial^2 \Delta G_{\mathrm{mix}}}{\partial x_{\mathrm{B}}^2}\right)_{T,P},
  \label{eq:gamma_main}
\end{equation}
where $x_{\mathrm B}$ is the atomic fraction of boron, with $\Gamma>0$ for a locally stable melt and $\Gamma<0$ inside the spinodal region.
The learned functional provides continuous predictions of $\Delta G_{\mathrm{mix}}$ and $\Gamma$ across composition, temperature and pressure, resolving the negative curvature inaccessible to equilibrium fluctuation measurements (\cref{fig:feb}a).
The AIPF model predicts a B-rich spinodal region at ambient pressure, and both increasing pressure and increasing temperature stabilize the melt (\cref{fig:feb}b,c).
At 10\,GPa, the FeB$_4$ composition exits the spinodal region, consistent with the proposed thermodynamic indicator of synthesizability.
Beyond agreeing with the spinodal classification from MD,
the AIPF model predicts the sign of the free-energy curvature for states whose stability remains unresolved in the MD trajectories (\cref{fig:feb}d).

\medskip
\noindent\textbf{Hydrogen--helium mixture.} 
To demonstrate applicability beyond synthesizability, we apply the framework to
hydrogen--helium (H--He) mixtures, which make up most of the mass of gas giants such as Jupiter and Saturn.
Inside gas giants, helium is hypothesized to become immiscible in metallic hydrogen and settle as droplets toward the planetary center, a process known as helium rain. 
Whether and where helium rain occurs depends on the immiscibility boundary as a function of helium fraction $x_{\mathrm{He}}$, pressure $P$ and temperature $T$, while its subsequent evolution depends on the kinetics of droplet nucleation, growth and transport.
Experiments at planetary interior conditions are extremely challenging, and direct AIMD simulations are limited in accessible length and time scales.  MLIP-driven MD simulations extend thermodynamic predictions to larger systems~\cite{chang2024,wangHydrogenheliumImmiscibilityBoundary2026a}, but these systems remain far smaller than the length and time scales of helium rain dynamics (\cref{fig:appendix_hhe_scaling}).
How helium droplets nucleate, grow and redistribute helium therefore remains an open question.

We train a single AIPF model, spanning 200--800\,GPa and 2000--12\,000\,K, on trajectories from molecular dynamics driven by a MACE potential~\cite{wangHydrogenheliumImmiscibilityBoundary2026a}.
The immiscibility boundary across $(x_{\mathrm{He}},P,T)$~(\cref{fig:hhe_main}a,b), together with the free energy of mixing and its curvature~(\cref{fig:appendix_hhe_free_energy}), follows directly from the learned functional.
We compare our predicted binodal and spinodal along the protosolar isopleth $x_{\mathrm{He}}=0.089$ with prior results (\cref{fig:hhe_main}d) and with the interior profiles of Jupiter and Saturn (\cref{fig:hhe_main}e).
At the protosolar composition, Saturn's interior profiles intersect the predicted immiscibility region, in agreement with previous studies~\cite{lorenzenMetallizationHydrogenheliumMixtures2011,chang2024,wangHydrogenheliumImmiscibilityBoundary2026a}.
Jupiter's profiles lie above but close to the immiscibility boundary. With more accurate MLIPs and planetary thermal profiles, our framework would have the potential to give a definitive answer on whether helium demixes in Jupiter.

Beyond predicting the immiscibility boundary, the learned AIPF model directly simulates how helium rain develops at mesoscopic scales.
We simulate helium rain for $1\,$ns in a column corresponding to 2.2 million atoms (\cref{fig:hhe_main}f,g), beyond the reach of AIMD or MLIP-driven MD at comparable computational cost
(\cref{fig:appendix_hhe_scaling})%
.
The simulation includes an enhanced gravitational potential absent from training%
.
The initially uniform density fields fluctuate within the metastable region until helium droplets nucleate and grow. Under gravity, helium accumulates in the lower region and is depleted from the upper region. As helium depletion drives the upper region across the immiscibility boundary, nucleation stops there and the rain zone shifts downward.
These results provide a dynamical picture of helium rain, from droplet formation in an initially homogeneous mixture to helium redistribution and the downward shift of the rain zone. They also demonstrate the scalability of the AIPF model and its ability to include an imposed external field.

\section*{Discussion}

We have developed a framework for learning \textit{ab initio} phase-field models that accurately predict mesoscopic thermodynamics and dynamics from microscopic data.
The framework captures the contrasting effects of pressure, which stabilizes the Fe--B melt but promotes demixing in the H--He mixture.
Our framework and neural cDFT~\cite{sammullerNeuralFunctionalTheory2023,buiLearningClassicalDensity2025,buiUnifiedMachineLearning2026,chengEquivariantLearningTransferable2026} both replace prescribed free-energy forms with flexible, learned functionals.
Neural cDFT primarily targets equilibrium thermodynamics, whereas we jointly learn the free-energy functional and mobility from dynamics, with equilibrium predictions obtained directly from the free-energy functional.
The training trajectories, starting from unstable homogeneous mixtures or slab configurations, naturally sample diverse density profiles as the systems relax toward equilibrium.
This diversity enables training on short trajectories without external potentials, which makes the framework data efficient compared with equilibrium sampling under a range of external potentials.
Learning from these trajectories also gives the AIPF model direct access to the spinodal region, which equilibrium approaches cannot sample and can reach only by extrapolation.

The accuracy of the learned AIPF model depends on the underlying MLIP, and advances in these potentials could make the microscopic data both more accurate and cheaper to generate.
The free-energy and mobility parametrizations used here represent one implementation of the AIPF framework. The framework can accommodate more general parametrizations, allowing greater predictive accuracy and broader applicability as more diverse training data become available. 
The present model resolves only density fields.
In the helium-rain simulation, helium therefore accumulates in the lower region through diffusion under enhanced gravity rather than through hydrodynamic droplet settling.
Incorporating momentum density would extend the framework to a broader range of fluid phenomena, capturing how flow shapes phase separation and transport.
The Markovian assumption may also become inadequate when memory persists on the time scale of the resolved fields, as can occur in supercooled liquids and glasses.
Learning memory kernels and correlated noise offers a possible extension, as has been done for collective variables on crystal lattice sites~\cite{xieAbInitioGeneralized2024} and for coarse-grained particles with many-body, state-dependent memory~\cite{lyuConstructionCoarseGrainedMolecular2023,geDataDrivenLearningGeneralized2024a}. 
Extending these approaches to continuum density fields requires memory kernels that are nonlocal in both space and time.
Beyond the fluids studied here, applications to solids will require training data that capture rare atomic rearrangements. 
Generating such data efficiently, for example through kinetic Monte Carlo with barriers computed from first principles, is a natural next step.

Looking ahead, we envision \textit{ab initio} phase-field modeling as an accurate and scalable tool for mesoscopic simulation grounded in microscopic theory and data. 
Training across elements, compositions and thermodynamic conditions could establish mesoscopic foundation models with reusable free-energy functionals and mobilities, analogous to CALPHAD (calculation of phase diagrams) databases~\cite{steinbachCALPHADPhaseFieldModeling2007} but built on general free-energy functionals rather than parametric bulk free energies.
Such mesoscopic foundation models could enable accurate thermodynamic and dynamical predictions in regimes where experiments are difficult and direct MD simulations are prohibitively costly.

\section*{Methods}
\noindent\textbf{Theoretical derivation.}
We outline the derivation here and provide the full details in Supplemental Material, Section S1.
We consider a multicomponent system containing $N_\alpha$ atoms of each species $\alpha$ in a simulation cell of volume $V$ with periodic boundary conditions.
The microscopic state $\mathbf Z$ comprises their positions $\mathbf r_j^\alpha$ and momenta $\mathbf p_j^\alpha$,
extended by thermostat variables where applicable.
We consider microscopic dynamics that is deterministic on this phase space and possesses an invariant distribution $\rho^{\mathrm{eq}}(\mathbf z)$, where $\mathbf z$ denotes a point of this phase space.
Molecular dynamics in the microcanonical (NVE) and canonical (NVT) ensembles falls within this class, provided the latter is generated by the Nos\'e--Hoover thermostat, whose dynamics is deterministic on the extended phase space and whose invariant measure is known~\cite{martynaNoseHooverChains1992}.
For any set of collective observables $\mathbf A(\mathbf Z(t))$ defined on the phase space, the Zwanzig projection operator maps a phase function $f(\mathbf z)$ to its conditional average at a fixed value $\mathbf a$ of the resolved variables,
\begin{equation*}
  \mathcal P_{\mathbf a}f
  =\frac{1}{\Omega(\mathbf a)}
   \int\!d\mathbf z\,\rho^{\mathrm{eq}}(\mathbf z)\,
      \delta\big(\mathbf A(\mathbf z)-\mathbf a\big)\,f(\mathbf z).
\end{equation*}
The normalization $\Omega(\mathbf a)$ is the marginal equilibrium distribution of the resolved variables,
\begin{equation*}
  \Omega(\mathbf a)=\int\!d\mathbf z\,\rho^{\mathrm{eq}}(\mathbf z)\,\delta\big(\mathbf A(\mathbf z)-\mathbf a\big),
\end{equation*}
and it defines the free energy of the resolved variables,
\begin{equation}
  \mathcal F(\mathbf a)=-k_{\mathrm B}T\ln\Omega(\mathbf a).
  \label{eq:freeenergy}
\end{equation}
For NVT dynamics, $\rho^{\mathrm{eq}}$ depends on the temperature, and every quantity defined through it carries a parametric dependence on the temperature, which we keep implicit for notational simplicity.
For NVE dynamics, $T$ in \cref{eq:freeenergy} is a fixed constant that drops out of the exact relations (Supplemental Material, Section S1).
Applying $\mathcal P_{\mathbf a}$ to the microscopic dynamics yields the GLE, which describes the nonequilibrium evolution of $\mathbf A(\mathbf Z(t))$~\cite{zwanzigMemoryEffectsIrreversible1961,grabert1982projection,schillingCoarsegrainedModellingOut2022}.
Specifically, $\mathbf A$ contains the independent components of the band-limited species density modes,
\begin{equation*}
  \hat{\boldsymbol\rho}=\big(\hat\rho_{\mathbf k,\alpha}\big)_{\lvert\mathbf k\rvert\le k_{\mathrm{cut}}},
  \qquad
  \hat\rho_{\mathbf k,\alpha}(\mathbf z)=\mathop{\textstyle\sum}\nolimits_{j=1}^{N_\alpha}e^{-i\mathbf k\cdot\mathbf r_j^{\alpha}}.
\end{equation*}
Here, $\hat\rho_{\mathbf k,\alpha}$ are the Fourier coefficients of the microscopic number density $\mathop{\textstyle\sum}\nolimits_j\delta(\mathbf r-\mathbf r_j^\alpha)$.
Throughout this paper, we use a hat $\hat{\cdot}$ to denote the spatial Fourier transform.
These modes define the band-limited density field in real space,
\begin{equation}
  \rho_\alpha(\mathbf r,t)=\frac1V\mathop{\textstyle\sum}\nolimits_{\lvert\mathbf k\rvert\le k_{\mathrm{cut}}}\hat\rho_{\mathbf k,\alpha}\big(\mathbf Z(t)\big)\,e^{i\mathbf k\cdot\mathbf r}.
  \label{eq:bandlimited}
\end{equation}%
The resulting GLE takes the form,
\iftwocol
\begin{equation}
\begin{split}
  \partial_t\hat{\boldsymbol\rho}
  &=\int_0^t\!ds\,\Big[
      k_{\mathrm B}T\,
      \frac{\partial}{\partial\hat{\boldsymbol\rho}}\!\cdot\!
      \mathbf K(\hat{\boldsymbol\rho}(t-s),s)\\
     &\quad-\mathbf K(\hat{\boldsymbol\rho}(t-s),s)\,
      \frac{\partial\mathcal F(\hat{\boldsymbol\rho}(t-s))}{\partial\hat{\boldsymbol\rho}}
    \Big]
   +\tilde{\mathbf R}(t),
\end{split}
  \label{eq:gle}
\end{equation}
\else
\begin{equation}
  \partial_t\hat{\boldsymbol\rho}
  =\int_0^t\!ds\,\Big[
      k_{\mathrm B}T\,
      \frac{\partial}{\partial\hat{\boldsymbol\rho}}\!\cdot\!
      \mathbf K(\hat{\boldsymbol\rho}(t-s),s)
      -\mathbf K(\hat{\boldsymbol\rho}(t-s),s)\,
      \frac{\partial\mathcal F(\hat{\boldsymbol\rho}(t-s))}{\partial\hat{\boldsymbol\rho}}
    \Big]
   +\tilde{\mathbf R}(t),
  \label{eq:gle}
\end{equation}
\fi
where $\mathbf K(\hat{\boldsymbol\rho}(t-s),s)$ is the memory kernel, which describes how the density modes at the current time are affected by the past configurations of the field.
$\mathbf K$ is a matrix on the mode indices, and the divergence is taken in its second index, $\big(\partial/\partial\hat{\boldsymbol\rho}\cdot \mathbf K\big)_\mu=\mathop{\textstyle\sum}\nolimits_\nu\partial K_{\mu\nu}/\partial\hat\rho_\nu$.
$\tilde{\mathbf R}(t)$ is the fluctuating force satisfying the fluctuation--dissipation relation, which fixes its covariance in terms of the memory kernel~\cite{hijonMoriZwanzigFormalism2010a,kuhnhold2019}.
The construction is exact whether the projection is defined with respect to the NVE or the NVT ensemble~\cite{kuhnhold2019,hijonMoriZwanzigFormalism2010a}.

The three assumptions introduced with \cref{eq:meso} act on \cref{eq:gle} as follows and are stated in their precise form 
as Assumptions 1 to 3 in Supplemental Material, Section S1\,C.
First, we assume that the memory kernel and its first derivatives with respect to the resolved state decay on a time $\tau_{\mathrm{mem}}$ much shorter than the evolution time $\tau_{\mathrm{field}}$ of every retained density mode.
Only time lags $s\lesssim\tau_{\mathrm{mem}}$ contribute appreciably to the memory integral, and during this interval the density changes negligibly, so $\hat{\boldsymbol\rho}(t-s)\approx\hat{\boldsymbol\rho}(t)$ and the kernel is replaced by $2\bar{\mathbf M}\,\delta(s)$, with $\bar{\mathbf M}(\hat{\boldsymbol\rho})=\int_0^\infty\!ds\,\mathbf K(\hat{\boldsymbol\rho},s)$ the integrated mobility.
The fluctuation--dissipation relation together with this delta approximation gives the covariance of $\tilde{\mathbf R}$ as $2k_{\mathrm B}T\bar{\mathbf M}\,\delta(s)$.
Second, we assume that the integrated mobility takes the form
\begin{equation*}
  \bar M_{\mathbf k\alpha,\mathbf k'\beta}
  =-(\mathbf k\cdot\mathbf k')\int_V
  M_{\alpha\beta}(\boldsymbol\rho(\mathbf r),T)
  e^{-i(\mathbf k+\mathbf k')\cdot\mathbf r}\,\mathrm d\mathbf r,
\end{equation*}
where $\mathbf M$ is a symmetric positive semidefinite matrix on the species indices, depending only on the local species densities and temperature.
Under this restriction the divergence of the mode-space mobility, $\partial/\partial\hat{\boldsymbol\rho}\cdot\bar{\mathbf M}$, vanishes, because the retained band contains conjugate wavevector pairs.
Third, we replace the exact force by a zero-mean white noise $\hat{\mathbf R}$, Gaussian conditional on the current state, with this covariance, which is obtained in real space as $\nabla\!\cdot\!\big[\sqrt{2k_{\mathrm B}T\mathbf M}\,\boldsymbol\xi\big]$.
The three assumptions then give $\partial_t\hat{\boldsymbol\rho}=-\bar{\mathbf M}\,\partial\mathcal F/\partial\hat{\boldsymbol\rho}+\hat{\mathbf R}$, which in real space is \cref{eq:meso}, and which admits the Boltzmann distribution $P[\boldsymbol\rho]\propto e^{-\mathcal F([\boldsymbol\rho],T)/k_{\mathrm B}T}$ as its invariant density.
\medskip

\medskip
\noindent\textbf{Training data.}
We first smooth the microscopic density of each species by replacing each $\delta(\mathbf r-\mathbf r_j^{\alpha})$ with a normalized Gaussian $G_\sigma(\mathbf r-\mathbf r_j^{\alpha})$, where $G_\sigma(\mathbf r)=(2\pi\sigma^2)^{-3/2}e^{-|\mathbf r|^2/(2\sigma^2)}$.
In Fourier space, this multiplies each mode $\hat\rho_{\mathbf k,\alpha}$ by $e^{-k^2\sigma^2/2}$.
We then retain modes with $|\mathbf k|\le k_{\mathrm{cut}}$ and transform back to real space, which gives the Gaussian-weighted band-limited field
\begin{equation}
  \rho_\alpha^\sigma(\mathbf r,t)=\frac1V\mathop{\textstyle\sum}\nolimits_{\lvert\mathbf k\rvert\le k_{\mathrm{cut}}}e^{-k^2\sigma^2/2}\,\hat\rho_{\mathbf k,\alpha}\big(\mathbf Z(t)\big)\,e^{i\mathbf k\cdot\mathbf r}.
  \label{eq:kde}
\end{equation}%
We use $\rho_\alpha^\sigma$ rather than the band-limited field $\rho_\alpha$ of \cref{eq:bandlimited} for two reasons.
First, truncating the Fourier series at $k_{\mathrm{cut}}$ produces oscillations in real space, so $\rho_\alpha$ can take unphysical negative values. We choose $k_{\mathrm{cut}}=4/\sigma$ because the Gaussian weight has already decayed to $e^{-8}$ there, so the cutoff has little effect and $\rho_\alpha^\sigma$ is essentially the sum of Gaussians, which is positive.
Second, the Gaussian smoothing makes the temporal evolution of the density field smoother and thus easier to learn.
For notational simplicity, we write $\rho_\alpha$ for $\rho_\alpha^\sigma$ in what follows.
We store the Fourier modes rather than the real-space density fields to reduce storage requirements.

\textit{Lennard-Jones.}
We adopted the symmetric binary Lennard-Jones mixture of Das \textit{et al.}~\cite{dasStaticDynamicCritical2006}, with $\epsilon_{AB}=0.5\,\epsilon_{AA}$ and a common diameter $\sigma_{AA}$. The pair potential was truncated and force-shifted at $2.5\,\sigma_{AA}$, and the total number density was $\rho_A+\rho_B=1.0\,\sigma_{AA}^{-3}$.
We used overdamped Langevin dynamics with translational damping $\gamma_t=2.0$ and a time step of $2\times10^{-4}\,\tau$, where $\tau=\sigma_{AA}\sqrt{m/\epsilon_{AA}}$ and $m$ is the common particle mass. For the Lennard-Jones mixture, we express length, temperature and time in units of $\sigma_{AA}$, $\epsilon_{AA}/k_{\mathrm B}$ and $\tau$, respectively.
Two run families supply different information.
Slabs of 3456 particles at $T=1.00$--$1.70$ provide the interface dynamics on which the drift term is fitted, with $T=1.20$ held out.
Homogeneous cubes of 4000 particles at $T=1.50$--$1.80$ supply the equilibrium anchors.
The coarse-graining width is $\sigma=1.5\,\sigma_{AA}$, giving $k_{\mathrm{cut}}=2.67\,\sigma_{AA}^{-1}$.

\textit{Iron--boron.}
We used a MACE-MH-1 potential with its r$^2$SCAN head, fine-tuned for the Fe--B melt~\cite{zhangInterfacialmeltStabilityThermodynamic2026}.
We generated training data in the NPT ensemble with a Nos\'e--Hoover thermostat and a Martyna--Tobias--Klein barostat at 0, 5 and $10\,$GPa.
The cell dimensions vary little along a trajectory, and positions are rescaled to the mean cell before the density modes are formed. The rescaled trajectories are then treated as fixed-cell data, to which the derivation above applies.
All Fe--B simulations used a time step of $1\,$fs.
Only cubes were used for training, since the coexisting compositions are unknown and a slab cannot be prepared.
Cubes of 3456 atoms covered $x_{\mathrm{B}}=0.10$--$0.98$ and $T=1200$--$2600\,$K, with run lengths of $50$ to $500\,$ps.
We measured the equilibrium number density $n(x_{\mathrm{B}},T;P)$ at every state point using isotropic NPT runs of 512 atoms, with $5\,$ps of melting, $5\,$ps of equilibration and $5\,$ps of averaging.
The coarse-graining width is $\sigma=2.0\,$\AA{}, giving $k_{\mathrm{cut}}=2.0\,$\AA$^{-1}$.

\textit{Hydrogen--helium.}
We used a MACE potential trained on vdW-DF reference calculations~\cite{wangHydrogenheliumImmiscibilityBoundary2026a}.
All simulations ran in the NPT ensemble with the same thermostat and barostat at 200, 400, 600 and $800\,$GPa, spanning 2000 to $12\,000\,$K.
Cubes of 3456 atoms covered $x_{\mathrm{He}}=0.05$--$0.98$, and slabs of 2000 to 4700 atoms each paired two compositions between pure hydrogen and pure helium (Supplemental Material, Section S3).
We used time steps of $0.1\,$fs for slabs and $0.2\,$fs for homogeneous cubes and equation-of-state calculations.
We measured the equilibrium number density $n(x_{\mathrm{He}},T;P)$ at every state point using short isotropic NPT runs of 512 atoms, with $1\,$ps of equilibration followed by $1\,$ps of averaging. The resulting table was used to set the density at which each cube was built. The small box suppresses phase separation, so the table gives the density of the homogeneous state even inside the miscibility gap.
The coarse-graining width is $\sigma=2.0\,$\AA{}, giving $k_{\mathrm{cut}}=2.0\,$\AA$^{-1}$.
Further details for each system are given in Supplemental Material, Section S3.

\medskip
\noindent\textbf{Model parametrization.}
We denote by $\theta$ all trainable parameters, namely the weights of the neural networks and scalar coefficients such as $\gamma$ and $E_a$ introduced below, and write the learned free energy and mobility as $\mathcal F_\theta$ and $\mathbf M_\theta$.
The learned free energy is a Helmholtz functional of the species densities and temperature, given by the sum of an intrinsic part and a coupling to an external potential,
\begin{equation}
  \mathcal F_\theta([\boldsymbol\rho],T)
  =\mathcal F_{\mathrm{int},\theta}([\boldsymbol\rho],T)
  +\int \mathbf V_{\mathrm{ext}}(\mathbf r)\!\cdot\!\boldsymbol\rho(\mathbf r)\,
   \mathrm d\mathbf r ,
  \label{eq:functional}
\end{equation}
and the intrinsic part splits into an ideal-gas term, known exactly, and an excess term that carries all interparticle correlations,
\begin{equation*}
\begin{split}
  \mathcal F_{\mathrm{int},\theta}&=\mathcal F_{\mathrm{id}}+\mathcal F_{\mathrm{exc},\theta},\\
  \mathcal F_{\mathrm{id}}&=k_{\mathrm B}T\mathop{\textstyle\sum}\nolimits_\alpha\int\rho_\alpha\big[\ln(\rho_\alpha\Lambda_\alpha^3)-1\big]\,\mathrm d\mathbf r ,
\end{split}
\end{equation*}
where $\Lambda_\alpha$ is the thermal wavelength.
In all three systems, $\mathcal F_{\mathrm{exc},\theta}$ takes the rung 3 form of \cref{eq:pair_b}, with a learned local free-energy density $f_{\mathrm{loc},\theta}(\boldsymbol\rho,T)$ and a learned nonlocal kernel $\mathbf W_\theta$ whose components $W_{\alpha\beta,\theta}(r)$ depend only on the separation $r=\lvert\mathbf r-\mathbf r'\rvert$.
The mobility depends only on the densities and temperature at the same point and is parametrized as
\begin{equation*}
  \mathbf M_\theta=\mathbf C_\theta^{1/2}\mathbf D^{1/2}\mathbf L_\theta\mathbf L_\theta^{\!\top}\mathbf D^{1/2}\mathbf C_\theta^{1/2},
\end{equation*}
where $\mathbf D(\boldsymbol\rho)=\mathrm{diag}(\boldsymbol\rho)$, the lower-triangular $\mathbf L_\theta(\boldsymbol\rho)$ is represented by a multilayer perceptron (MLP) taking the densities as inputs, and $\mathbf C_\theta(T)=\mathrm{diag}\big(c_{\alpha,\theta}(T)\big)$ carries the temperature dependence.
This construction ensures that $\mathbf M_\theta$ is symmetric and positive semidefinite.

We begin with stronger physical priors and adopt more general parametrizations only when the simpler forms are insufficient to describe the system.
For $f_{\mathrm{loc},\theta}$, we first consider a temperature expansion,
\begin{equation*}
  f_{\mathrm{loc},\theta}(\boldsymbol\rho,T)
  =\mathop{\textstyle\sum}\nolimits_i u_{i,\theta}(\boldsymbol\rho)b_i(T),
\end{equation*}
using simple basis functions such as $1$, $T$, $T\ln T$ and $T^2$.
The linear form suffices for the Lennard-Jones mixture, whereas H--He requires an additional $T\ln T$ term.
When these expansions are insufficient, as for the strongly asymmetric Fe--B melt, we instead adopt a general MLP for $f_{\mathrm{loc},\theta}(\boldsymbol\rho,T)$.
We initially take the mobility to be temperature independent, with $\mathbf C_\theta=\mathbf I$, and introduce Arrhenius factors when the mobility anchors show temperature dependence, as for Fe--B and the Lennard-Jones mixture.

\textit{Lennard-Jones.}
The Lennard-Jones mixture is nearly incompressible, so we take the local total density $\rho_A+\rho_B$ to be $1.0\,\sigma_{AA}^{-3}$ at every point.
The dependence of the free energy and mobility on the species densities therefore reduces to a dependence on the composition field $x_A=\rho_A/(\rho_A+\rho_B)$, giving $\mathcal F([x_A],T)$ and $M(x_A,T)$.
The ideal part of the free energy, the local free-energy density and the mobility are
\begin{equation*}
\begin{aligned}
  f_{\mathrm{id}}&=T\big[x_A\ln x_A+(1-x_A)\ln(1-x_A)\big],\\
  f_{\mathrm{loc},\theta}&=u_{0,\theta}(x_A)+T\,u_{1,\theta}(x_A),\\
  M_\theta&=\gamma\,x_A(1-x_A)\,c_\theta(T),\qquad c_\theta(T)\propto\exp(-E_a/T),
\end{aligned}
\end{equation*}
with $\mathcal F_{\mathrm{id}}=\int f_{\mathrm{id}}\,\mathrm d\mathbf r$ and $f_{\mathrm{id}}$ defined up to a constant.
Here, $u_{0,\theta}$ is an even Taylor polynomial in $x_A-\tfrac12$ up to eighth order and $u_{1,\theta}$ is an input-convex neural network. In the pair term, $\boldsymbol\rho$ is replaced by the scalar field $x_A-\tfrac12$ and $\mathbf W_\theta$ by a scalar radial kernel $W_\theta$ of finite range $R_{\mathrm{cut}}=3.0\,\sigma_{AA}$. The mobility is the scalar form of $\mathbf M_\theta$, with $x_A(1-x_A)$ in place of $\mathbf D$ and a learned constant $\gamma$ in place of $\mathbf L_\theta\mathbf L_\theta^{\!\top}$. Both $\gamma$ and $E_a$ are learned.
For $u_{0,\theta}$, we found that the polynomial fits this single-variable function better than an MLP. The entire expression is even in $x_A-\tfrac12$ because of the $A$--$B$ symmetry.

\textit{Baselines.}
For the Lennard-Jones mixture, we compare with two rung 1 forms (\cref{fig:ladders}b), trained with the same data, loss, anchors and mobility form as the AIPF model.
The Flory--Huggins baseline keeps $\mathcal F_{\mathrm{id}}$ and takes
\begin{equation}
  \mathcal F_{\mathrm{exc},\theta}=\int\Big[w\,x_A(1-x_A)+\tfrac{\kappa}{2}\lvert\nabla x_A\rvert^2\Big]\,\mathrm d\mathbf r,
  \label{eq:fh}
\end{equation}
with temperature-independent $w$ and $\kappa$, both learned.
The Landau baseline drops the ideal term and takes
\begin{equation}
\begin{aligned}
  \mathcal F_{\mathrm{int},\theta}&=\int\Big[\tfrac{a}{2}(T-T_{\mathrm c})(x_A-\tfrac12)^2\colbreak
  {}+b\,(x_A-\tfrac12)^4+\tfrac{\kappa}{2}\lvert\nabla x_A\rvert^2\Big]\,\mathrm d\mathbf r,
\end{aligned}
  \label{eq:landau}
\end{equation}
with $a,b>0$, $T_{\mathrm c}$ and $\kappa$ learned.
Without the ideal term, it does not confine $x_A$ to $[0,1]$.

\textit{Iron--boron.}
An MLP represents the density and temperature dependence of $f_{\mathrm{loc},\theta}$.
The kernel $\mathbf W_\theta$ is a symmetric $2\times2$ matrix of radial functions with cutoff $R_{\mathrm{cut}}=4.5\,$\AA{}.
We learn one activation energy $E_\alpha$ per species for the mobility.

\textit{Hydrogen--helium.}
We use the temperature expansion
\begin{equation*}
  f_{\mathrm{loc},\theta}=u_{0,\theta}(\boldsymbol\rho)+k_{\mathrm B}T\,u_{1,\theta}(\boldsymbol\rho)+k_{\mathrm B}T\ln(T/T_{\mathrm{ref}})\,u_{2,\theta}(\boldsymbol\rho),
\end{equation*}
where each $u_{i,\theta}$ is an MLP of the species densities and $T_{\mathrm{ref}}$ is a reference temperature.
The kernel $\mathbf W_\theta$ has the same parametrization and cutoff as for Fe--B.
The mobility has no explicit temperature dependence, with $\mathbf C_\theta=\mathbf I$.

\medskip
\noindent\textbf{Thermodynamic predictions.}
Evaluating the learned functional $\mathcal F_\theta$ in the absence of an external potential on a spatially uniform profile $\boldsymbol\rho(\mathbf r)=\bar{\boldsymbol\rho}$ and dividing by the system volume $V$ gives the bulk Helmholtz free-energy density $f(\bar{\boldsymbol\rho},T)$. 
For the Lennard-Jones mixture, we construct the phase diagram from $f$ at fixed total density $n=\mathop{\textstyle\sum}\nolimits_\alpha\bar\rho_\alpha$.
At each temperature, a common tangent to $f(x_A,T;n)$ gives the binodal compositions, while $(\partial^2 f/\partial x_A^2)_{T,n}<0$ defines the spinodal region, where the homogeneous mixture is unstable to infinitesimal composition fluctuations (\cref{fig:appendix_lj_free_energy}).
Between the binodal and the spinodal region, the mixture is metastable and phase separation requires nucleation.

For Fe--B and H--He, we instead construct phase diagrams at prescribed temperature and pressure, allowing the total density to vary with composition.
The pressure of a homogeneous state follows from the Euler relation,
\begin{equation}
  P(\bar{\boldsymbol\rho},T) = \boldsymbol\mu\!\cdot\!\bar{\boldsymbol\rho} - f ,
  \qquad
  \boldsymbol\mu = \frac{\partial f}{\partial\bar{\boldsymbol\rho}} .
  \label{eq:euler}
\end{equation}
Let $x_\alpha=\bar\rho_\alpha/\mathop{\textstyle\sum}\nolimits_\beta\bar\rho_\beta$ be the number fraction of species $\alpha$. For a binary mixture, we denote the fraction of the selected species by $x$, species $A$ for the Lennard-Jones mixture, boron for Fe--B and helium for H--He. 
The composition fixes the direction of $\bar{\boldsymbol\rho}$, since $\bar{\boldsymbol\rho}=n\mathbf e$ with $e_\alpha=x_\alpha$. At a specified temperature, the total number density $n$ is obtained by solving \cref{eq:euler} at the prescribed pressure and composition. The species densities then follow as $\bar\rho_\alpha=nx_\alpha$.
For the H--He mixture, \cref{fig:appendix_hhe_eos} compares the densities so obtained with the MD equation-of-state table.
At a prescribed pressure $P$, the Gibbs free energy per particle also follows from \cref{eq:euler},
\begin{equation*}
  g(x,T;P) = \frac{f+P}{n} = \mathop{\textstyle\sum}\nolimits_\alpha x_\alpha\mu_\alpha ,
\end{equation*}
and the Gibbs free energy of mixing per particle is
\begin{equation}
\begin{split}
  &\Delta G_{\mathrm{mix}}(x,T;P) = g(x,T;P) \colbreak
    {}- \big[(1-x)\,g(0,T;P) + x\,g(1,T;P)\big] .
\end{split}
  \label{eq:dgmix}
\end{equation}
At fixed $T$ and $P$, a common tangent to $\Delta G_{\mathrm{mix}}$ gives the binodal compositions, while $(\partial^2\Delta G_{\mathrm{mix}}/\partial x^2)_{T,P}<0$ defines the spinodal region. The metastable region lies between the binodal and the spinodal region, and their boundaries meet at the critical point. \Cref{fig:appendix_feb_dgmix,fig:appendix_hhe_free_energy} show $\Delta G_{\mathrm{mix}}$ and its curvature along the isobars of the Fe--B melt and the H--He mixture.
The thermodynamic factor of \cref{eq:gamma_main} is the reduced composition curvature of the Gibbs free energy per particle, taken at fixed $T$ and $P$ and written here for a general composition $x$. The learned functional gives $\Gamma$ at any state, and it is negative inside the spinodal region.
The equilibrium density under an external potential is defined in Supplemental Material, Section S2\,B.

\medskip
\noindent\textbf{Dynamical predictions.}
We integrate the learned AIPF model using a pseudospectral method on a periodic grid.
For the rung 3 functional (\cref{eq:pair_b}), the Fourier transform of the excess chemical potential $\delta\mathcal F_{\mathrm{exc},\theta}/\delta\boldsymbol\rho$ is $\widehat{\partial_{\boldsymbol\rho}f_{\mathrm{loc},\theta}}(\mathbf k)+\widehat{\mathbf W}_\theta(\mathbf k)\hat{\boldsymbol\rho}(\mathbf k)$.
The nonlocal convolution thus reduces to a matrix--vector product at each wavevector, which is computationally efficient in Fourier space.
The local, ideal-gas and external-potential contributions to the chemical potential are evaluated point by point in real space and then transformed to Fourier space using fast Fourier transforms.
We use a first-order implicit--explicit scheme, treating the dynamics linearized about the mean density implicitly, at the next time step, and the nonlinear remainder explicitly, at the current time step.
This splitting improves numerical stability, allowing larger time steps than a fully explicit scheme.
Thermal noise is added with the covariance set by the fluctuation--dissipation relation, using the same $\mathbf M_\theta$ as in the deterministic part.
The solver runs on a single NVIDIA A100 40\,GB GPU, with computational cost scaling approximately linearly with system size (\cref{fig:appendix_hhe_scaling}).
For helium-rain dynamics, the model simulates 1\,ns at a scale corresponding to 2.2 million atoms in about one day.
Implementation details are provided in Supplemental Material, Section S2\,A.

\medskip
\noindent\textbf{Fluctuation-based measurements.}
The structure factors and mobility can be measured from homogeneous equilibrium trajectories. 
The partial structure factors are
\begin{equation*}
  S_{\alpha\beta}(\mathbf k)=\frac{1}{\sqrt{N_\alpha N_\beta}}\big\langle\hat\rho_{\mathbf k,\alpha}\,\hat\rho_{\mathbf k,\beta}^{*}\big\rangle ,
\end{equation*}
where $\hat\rho_{\mathbf k,\alpha}$ for $\mathbf k\neq0$ are the density modes defined above, evaluated for NPT trajectories with positions rescaled componentwise as $\mathbf r_j'(t)=\mathbf r_j(t)\langle l\rangle/l(t)$, with $l(t)$ the instantaneous box length and $\langle l\rangle$ its time average. The wavevectors are $\mathbf k=2\pi\mathbf n/\langle l\rangle$, with integer $\mathbf n$. We average over frames and over wavevectors of equal magnitude $k=|\mathbf k|$, so the quantities below depend on $k$ only.
The Hessian of the free energy is related to the partial structure factors through 
\begin{equation}
  S_{\alpha\beta}(k)=\frac{k_{\mathrm B}T}{\sqrt{\bar\rho_\alpha\bar\rho_\beta}}\,\big[\mathbf H(k)^{-1}\big]_{\alpha\beta},\quad
  \mathbf H(k)=\widehat{\frac{\delta^2\mathcal F}{\delta\boldsymbol\rho\,\delta\boldsymbol\rho}}\bigg|_{\bar{\boldsymbol\rho}} .
  \label{eq:fdt_static}
\end{equation}
Here, $\mathbf H(k)$ is the Fourier transform of the second functional derivative at the uniform state, with $\mathbf H(0)=\partial^2 f/\partial\bar{\boldsymbol\rho}\,\partial\bar{\boldsymbol\rho}$, where $f$ is the bulk Helmholtz free-energy density.

The concentration structure factor is
\begin{equation*}
  S_{cc}(k)=x_\alpha x_\beta\big[x_\beta S_{\alpha\alpha}(k)+x_\alpha S_{\beta\beta}(k)-2\sqrt{x_\alpha x_\beta}\,S_{\alpha\beta}(k)\big].
\end{equation*}
We estimate its long-wavelength limit $S_{cc}(0)$ by fitting the Ornstein--Zernike form $a/(1+bk^2)$ to the low-$k$ shells of $S_{cc}(k)$ and taking the intercept $a$.
For a stable homogeneous state, the thermodynamic factor $\Gamma$ of \cref{eq:gamma_main} is fixed by the equilibrium concentration fluctuations,
\begin{equation*}
  \Gamma_{\mathrm{MD}}(x,T;P) = \frac{x(1-x)}{S_{cc}(0;x,T,P)} ,
\end{equation*}
where $x$ is the number fraction of the selected species. Only in a stable homogeneous state does this estimate coincide with the free-energy curvature (\cref{eq:gamma_main}). $\Gamma_{\mathrm{MD}}$ is nonnegative by construction. Inside the spinodal region, phase separation enhances long-wavelength concentration fluctuations, driving $\Gamma_{\mathrm{MD}}$ toward zero as the simulation time and system size increase. Within the accessible MD time scales it only approaches zero from above, whereas the functional returns the negative curvature directly. This is the difference displayed in \cref{fig:feb}d, and the reason the classification of \cref{fig:feb}a thresholds $\Gamma_{\mathrm{MD}}$ instead of testing its sign.

We obtain the mobility from the covariance of changes in the density modes over intervals $\Delta t$ longer than the microscopic memory time but shorter than the relaxation time of the resolved modes. The fluctuation--dissipation relation in \cref{eq:meso} gives, to leading order in $\Delta t$,
\begin{equation}
  \big\langle\Delta\hat\rho_{\mathbf k,\alpha}\,\Delta\hat\rho_{\mathbf k,\beta}^{*}\big\rangle
  = 2k_{\mathrm B}T\,k^{2}\langle V\rangle\,M_{\alpha\beta}\,\Delta t ,
  \label{eq:fdt_mobility}
\end{equation}
where $\Delta\hat\rho_{\mathbf k,\alpha}=\hat\rho_{\mathbf k,\alpha}\big(\mathbf Z(t+\Delta t)\big)-\hat\rho_{\mathbf k,\alpha}\big(\mathbf Z(t)\big)$ is the change in a density mode over the interval $\Delta t$ and $\langle V\rangle$ is the mean box volume. For H--He and Fe--B, we estimate $\mathbf M$ from the zero-lag intercept of a linear fit to finite-lag mobility estimates within this mesoscopic time window. For the Lennard-Jones mixture, we evaluate it directly at a lag of $\tau$ (Supplemental Material, Section S3).

\medskip

\noindent\textbf{Training objective.}
We learn the AIPF model by minimizing a loss function that compares its dynamics and thermodynamics with microscopic data.
The dynamical term compares the measured density evolution with the deterministic part of \cref{eq:meso},
\begin{equation*}
\begin{aligned}
  \mathcal L_{\mathrm{dyn}}&=\Big\langle\big\|\partial_t\boldsymbol\rho-\nabla\!\cdot\!\big[\mathbf M_\theta\nabla\boldsymbol\mu_\theta\big]\big\|_{H^{-1}}^2\Big\rangle,\\
  \|\mathbf f\|_{H^{-1}}^2&=\frac{1}{N_k}\mathop{\textstyle\sum}\nolimits_{0<|\mathbf k|\le k_{\mathrm{cut}}}\frac{|\hat{\mathbf f}(\mathbf k)|^2}{|\mathbf k|^2},
\end{aligned}
\end{equation*}%
where $\boldsymbol\mu_\theta(\mathbf r)=\delta\mathcal F_\theta/\delta\boldsymbol\rho(\mathbf r)$ are the chemical potentials of the current field, which become spatially constant at equilibrium (\cref{eq:el}), $\partial_t\boldsymbol\rho$ is estimated from the coarse-grained density fields over each trajectory segment, $\langle\cdot\rangle$ averages over these segments, and $N_k$ is the number of retained modes.
The factor $1/|\mathbf k|^2$ gives larger weight to the long-wavelength modes, which carry the slow mesoscopic dynamics.

The dynamical loss alone does not fully determine the AIPF model, for two reasons.
First, $\mathcal L_{\mathrm{dyn}}$ matches only the deterministic part of \cref{eq:meso}. The noise, which alone would fix the mobility, is averaged out. In the deterministic part, $\mathbf M_\theta$ and $\boldsymbol\mu_\theta$ enter only through the product $\mathbf M_\theta\nabla\boldsymbol\mu_\theta$, so the rescaling $(\boldsymbol\mu_\theta,\mathbf M_\theta)\rightarrow(\lambda\boldsymbol\mu_\theta,\mathbf M_\theta/\lambda)$ leaves $\mathcal L_{\mathrm{dyn}}$ unchanged.
Second, near the critical temperature the mixture undergoes critical slowing down.
As the composition curvature approaches zero, the restoring force weakens relative to thermal noise. The drift therefore becomes less informative about the free-energy curvature, making equilibrium fluctuation measurements particularly important near $T_{\mathrm c}$.
Both limitations are addressed with the mobility and structure factors measured from homogeneous runs.

The mobility anchor $\mathcal L_M$ removes the scale ambiguity through the fluctuation--dissipation relation, which connects the leading-order covariance of short-time density changes to $\mathbf M_\theta$. It penalizes the mean squared relative deviation between $\mathbf M_\theta$ and the mobility measured through \cref{eq:fdt_mobility}.
The static anchors $\mathcal L_S$ and $\mathcal L_{\mathrm{bulk}}$ supply the thermodynamic information by matching the Hessian of $\mathcal F_\theta$ to the measured structure factors through \cref{eq:fdt_static}. $\mathcal L_S$ compares the $2\times2$ matrices on the measured shells $k\neq0$, and $\mathcal L_{\mathrm{bulk}}$ compares the scalar $S_{cc}(0)$ from the Ornstein--Zernike intercept at $k=0$.
\Cref{fig:appendix_hhe_anchors} compares the AIPF model with the measurements at the H--He anchor states.

For Fe--B and H--He, simulated in the NPT ensemble, we add an isobar anchor $\mathcal L_P$ that constrains the equation of state. It compares the pressure obtained from $\mathcal F_\theta$ through \cref{eq:euler} with the prescribed pressure at each state point, using a mean squared normalized residual.
These data-fitting terms form the weighted sum
\begin{equation}
  \mathcal L=\mathcal L_{\mathrm{dyn}}
    +\lambda_M\,\mathcal L_M+\lambda_S\,\mathcal L_S
    +\lambda_{\mathrm{bulk}}\,\mathcal L_{\mathrm{bulk}}
    +\lambda_P\,\mathcal L_P ,
  \label{eq:loss}
\end{equation}
with $\lambda_S=\lambda_P=0$ for the Lennard-Jones mixture.
We add regularization terms for Fe--B and H--He to \cref{eq:loss}. The regularization terms and the weight of every term are given per system in Supplemental Material, Section S3.
\medskip

\section*{Data availability}
The trained AIPF models, the data underlying the figures and the pipeline that generates the molecular-dynamics trajectories and training datasets are available at \url{https://github.com/mengyi-chen/ab-initio-phase-field-models}.
The trajectories and datasets themselves are available from the corresponding author on reasonable request, owing to their size.
The MLIPs for hydrogen--helium and iron--boron were taken from Wang \textit{et al.}~\cite{wangHydrogenheliumImmiscibilityBoundary2026a} and Zhang \textit{et al.}~\cite{zhangInterfacialmeltStabilityThermodynamic2026}, respectively.

\section*{Code availability}
The code for training and simulating AIPF models is available at \url{https://github.com/mengyi-chen/ab-initio-phase-field-models}.

\section*{Acknowledgments}
This research is supported by the National Research Foundation, Singapore under its AI Singapore Programme (AISG Award No: AISG3-RP-2022-028). 
This research is supported by computational resources from the NUS High Performance Computing facility (NUSREC-HPC-00001).
The computational work for this article was partially performed on resources of the National Supercomputing Centre, Singapore (https://www.nscc.sg).

\section*{Author contributions}
Q.L. and M.C. conceived the study. M.C. carried out the research. P.Z. contributed to the research design and interpretation of the results. Z.Z. developed the MLIP for the iron--boron melt. All authors contributed to discussions and to writing the manuscript.

\section*{Competing interests}
The authors declare no competing interests.

\bibliographystyle{apsrev4-2}
\bibliography{references_arxiv}
\clearpage
\onecolumngrid
\appendix
\section{Additional figures}
\setcounter{figure}{0}
\renewcommand{\thefigure}{A\arabic{figure}}
\renewcommand{\theHfigure}{A\arabic{figure}}
\renewcommand{\appfigbegin}{\begin{center}}\renewcommand{\appfigend}{\end{center}\clearpage}\renewcommand{\appcaption}{\captionof{figure}}

\appfigbegin
    \centering
    \begin{minipage}{\linewidth}\centering
    \panelw{0.921\linewidth}{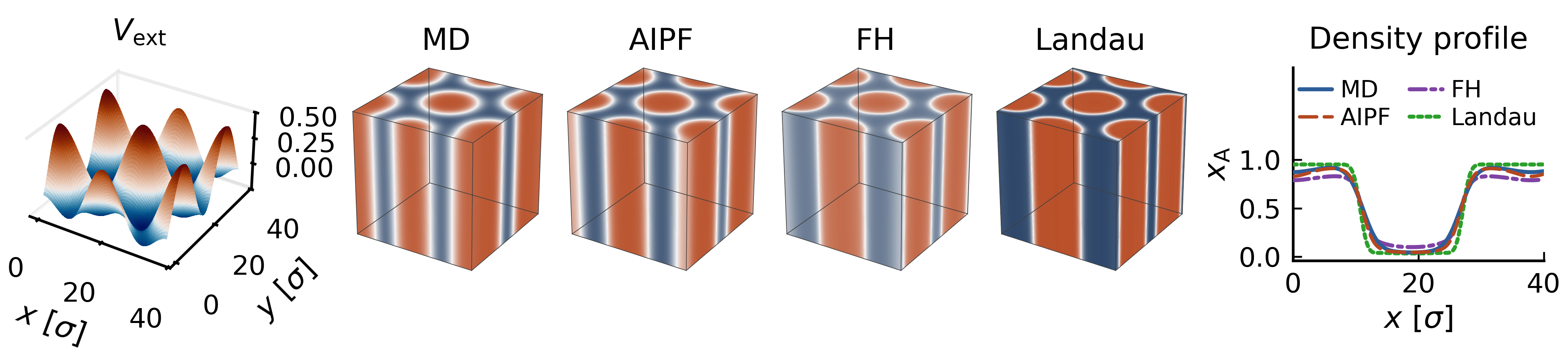}{a}\\[2pt]
    \panelw{0.921\linewidth}{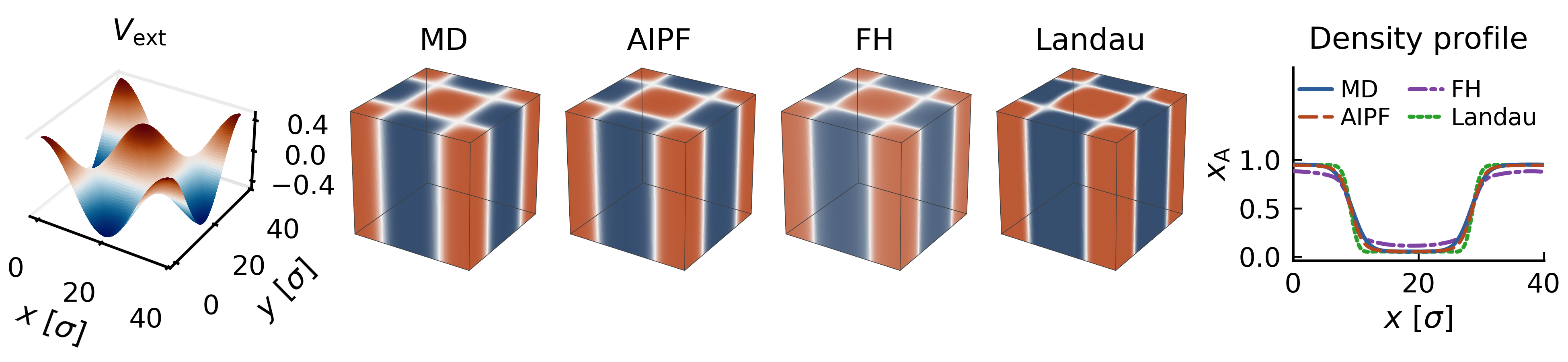}{b}
    \end{minipage}
    \appcaption{\textbf{Equilibrium distribution of the Lennard-Jones mixture under external potentials.}
    \textbf{a,b}, Equilibrium profiles are compared for hexagonal (\textbf{a}) and checkerboard (\textbf{b}) potentials at $T=1.20$.
    Each row shows, from left to right, the imposed potential $\mathbf V_{\mathrm{ext}}$, with its magnitude indicated by the vertical axis, the coarse-grained composition field from molecular dynamics, the fields of the AIPF, Flory--Huggins (FH) and Landau models, and profiles from all four along $y=L_y/2$.
    The fields of all three models are time averages of stochastic simulations started from a uniform composition.
    The external potentials and this temperature are not included in training.
    \label{fig:appendix_lj_vext}}
\appfigend

\appfigbegin
    \centering
    \setlength{\figlw}{\linewidth}%
    \begin{minipage}{\linewidth}\centering
    \makebox[0.921\figlw][s]{%
    \panelw{0.4955\figlw}{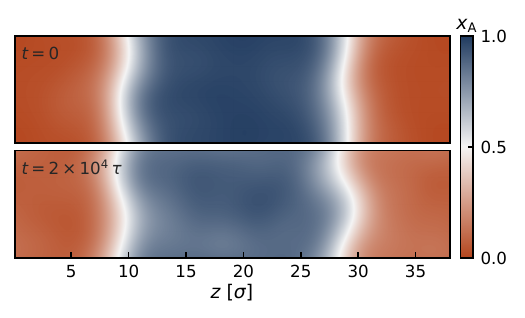}{a}\hfill
    \panelw{0.4062\figlw}{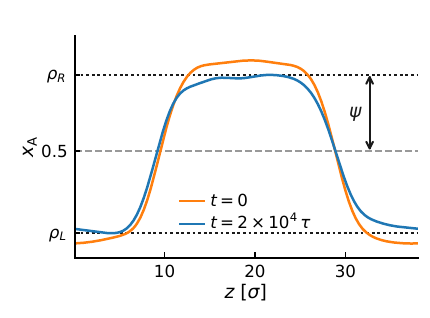}{b}}\\[4pt]
    \panelw{0.921\figlw}{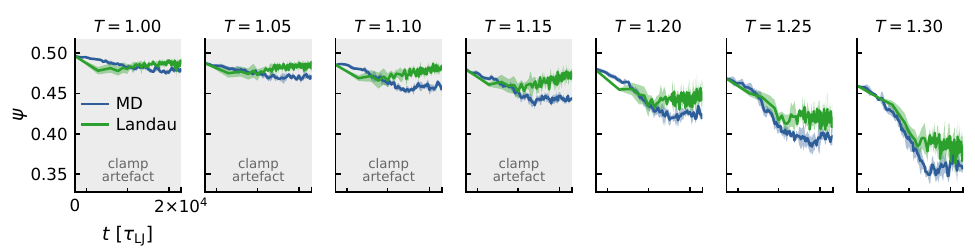}{c}\\[4pt]
    \panelw{0.921\figlw}{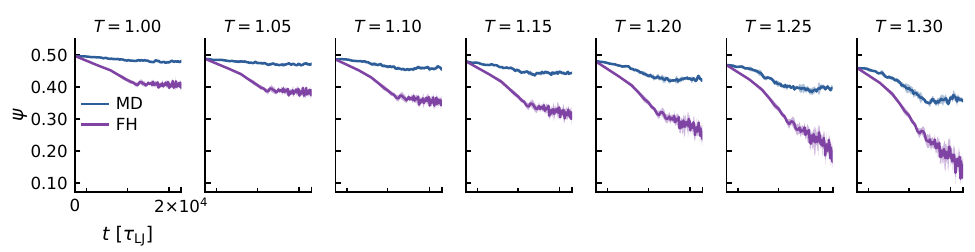}{d}\\[4pt]
    \panelw{0.921\figlw}{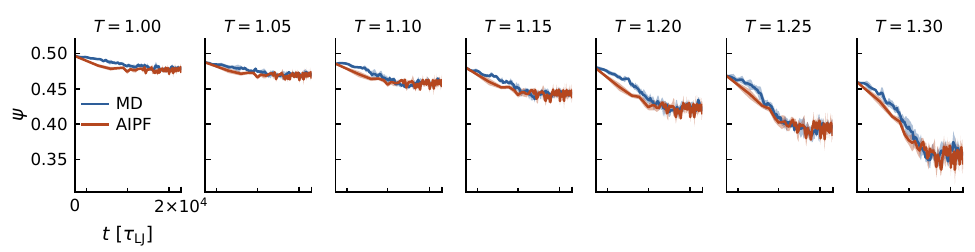}{e}
    \end{minipage}
    \appcaption{\textbf{Interface relaxation in the Lennard-Jones mixture.}
    \textbf{a,b}, MD composition fields (\textbf{a}) and corresponding $x$-averaged profiles (\textbf{b}) at $t=0$ and $t=2\times10^4\,\tau$ in a $9.52\times9.52\times38.10\,\sigma_{AA}^3$ box, with $z$ horizontal.
    The temperature, $T=1.20$, is not included in training.
    The composition contrast is defined as $\psi(t)=[x_R(t)-x_L(t)]/2$, where $x_R$ and $x_L$ are the upper and lower fitted plateaus of $x_A$ (right and left in \textbf{b}).
    \textbf{c--e}, Comparison of $\psi(t)$ with MD at seven temperatures for the Landau (\textbf{c}), Flory--Huggins (\textbf{d}) and AIPF (\textbf{e}) models.
    Curves show means over four independent seeds for MD and each model, and bands show one standard deviation.
    In the stochastic model simulations, $x_A$ is clamped to $[10^{-4},\,1-10^{-4}]$ at every step and the mean composition is restored.
    Below $T\approx1.17$, the Landau model predicts coexistence compositions outside $[0,1]$.
    In these gray panels of \textbf{c}, the subsequent increase in $\psi$ is a clamping artifact rather than physical interface relaxation.
    \label{fig:appendix_lj_slab}}
\appfigend

\appfigbegin
    \centering
    \begin{minipage}{\linewidth}\centering
    \includegraphics[width=0.743\linewidth]{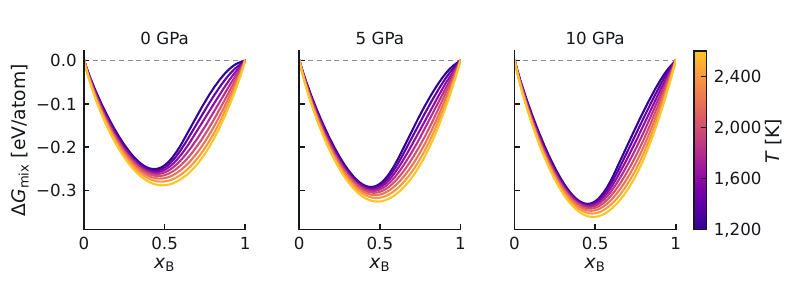}
    \end{minipage}
    \appcaption{\textbf{Mixing free energy of the iron--boron melt.}
    The Gibbs free energy of mixing per atom, $\Delta G_{\mathrm{mix}}$, is shown as a function of boron fraction $x_{\mathrm B}$ along the learned isobars at $0$, $5$ and $10\,$GPa. Colors indicate temperatures from $1200$ to $2600\,$K in increments of $200\,$K. The corresponding spinodal regions are identified by the thermodynamic factor $\Gamma$ in \cref{fig:feb}a,c.
    \label{fig:appendix_feb_dgmix}}
\appfigend

\appfigbegin
    \centering
    \begin{minipage}{\linewidth}\centering
    \begin{tikzpicture}
      \node[anchor=south west,inner sep=0] (im) {\includegraphics[width=89mm]{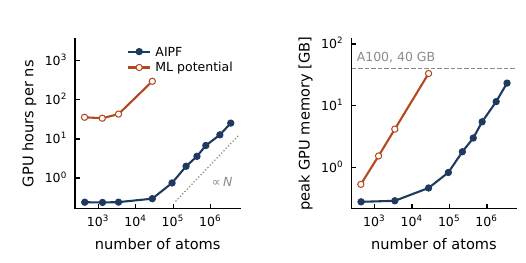}};
      \node[anchor=north west,font=\bfseries,inner sep=1pt] at (im.north west) {a};
      \node[anchor=north west,font=\bfseries,inner sep=1pt] at ([xshift=46.8mm]im.north west) {b};
    \end{tikzpicture}
    \end{minipage}
    \appcaption{\textbf{Comparison of computational scaling.}
\textbf{a}, GPU hours per simulated nanosecond and \textbf{b}, peak GPU memory as functions of the number of atoms for cubic boxes of the hydrogen--helium mixture at $x_{\mathrm{He}}=0.50$, $400\,$GPa and $9000\,$K.
All runs use a single NVIDIA A100 GPU with $40\,$GB of memory.
Molecular dynamics uses a MACE potential trained specifically for hydrogen--helium on vdW-DF data~\cite{wangHydrogenheliumImmiscibilityBoundary2026a}, with approximately $2\times10^{5}$ parameters.
Filled symbols denote the AIPF model simulated with a $5\,$fs time step and a grid spacing of approximately $0.7\,$\AA{}.
Open symbols, labeled ML potential, denote MD with a $0.2\,$fs time step.
The dotted line indicates linear scaling with system size.
\label{fig:appendix_hhe_scaling}}
\appfigend

\appfigbegin
    \centering
    \begin{minipage}{\linewidth}\centering
    \panelw{0.921\linewidth}{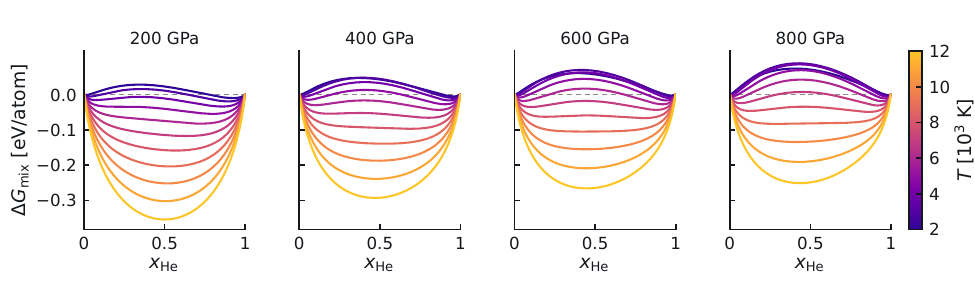}{a}\\[2pt]
    \panelw{0.921\linewidth}{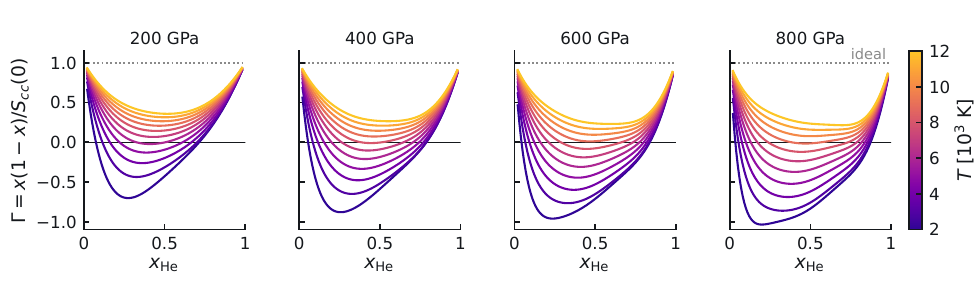}{b}
    \end{minipage}
    \appcaption{\textbf{Mixing free energy and thermodynamic factor of hydrogen--helium.}
    \textbf{a}, The Gibbs free energy of mixing per atom, $\Delta G_{\mathrm{mix}}$, is shown as a function of helium fraction along the learned isobars at $200$, $400$, $600$ and $800\,$GPa. Colors indicate temperatures from $2000$ to $12\,000\,$K in increments of $1000\,$K. Common tangents determine the binodal compositions.
    \textbf{b}, The corresponding thermodynamic factor $\Gamma$ (\cref{eq:gamma_main}). The dotted line marks ideal mixing ($\Gamma=1$), and the solid line marks the spinodal boundary ($\Gamma=0$). Negative values identify the spinodal region.
    \label{fig:appendix_hhe_free_energy}}
\appfigend

\appfigbegin
    \centering
    \begin{minipage}{\linewidth}\centering
    \panelw{0.290\linewidth}{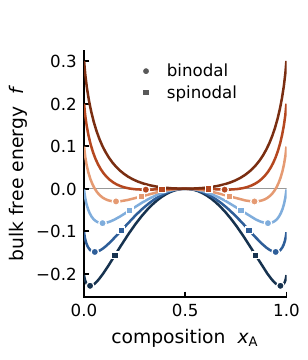}{a}\hspace{0.04\linewidth}
    \panelw{0.347\linewidth}{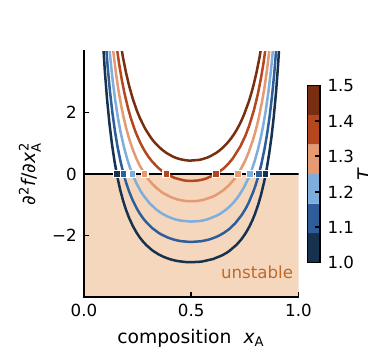}{b}
    \end{minipage}
    \appcaption{\textbf{Learned bulk free energy of the Lennard-Jones mixture.}
    \textbf{a}, The learned bulk free energy $f(x_A,T)$ is shown with the binodal and spinodal compositions, where $x_A$ is the number fraction of species $A$.
    \textbf{b}, The composition curvature $\partial^2 f/\partial x_A^2$ is negative in the spinodal region.
    \label{fig:appendix_lj_free_energy}}
\appfigend

\appfigbegin
    \centering
    \begin{minipage}{\linewidth}\centering
    \includegraphics[width=0.921\linewidth]{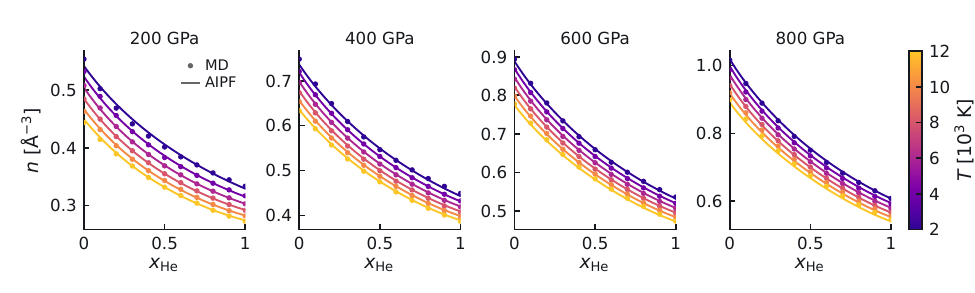}
    \end{minipage}
    \appcaption{\textbf{Comparison of the hydrogen--helium equation of state from the AIPF model and molecular dynamics.}
    Number density $n$ is shown as a function of helium fraction $x_{\mathrm{He}}$ at $200$, $400$, $600$ and $800\,$GPa. Colors indicate six isotherms from $2000$ to $12\,000\,$K in increments of $2000\,$K. Points denote molecular-dynamics reference states obtained with the machine-learning interatomic potential. Lines show the predictions of the AIPF model, obtained by solving the Euler relation (\cref{eq:euler}) at each composition and temperature.
    \label{fig:appendix_hhe_eos}}
\appfigend

\appfigbegin
    \centering
    \begin{minipage}{\linewidth}\centering
    \includegraphics[width=0.921\linewidth]{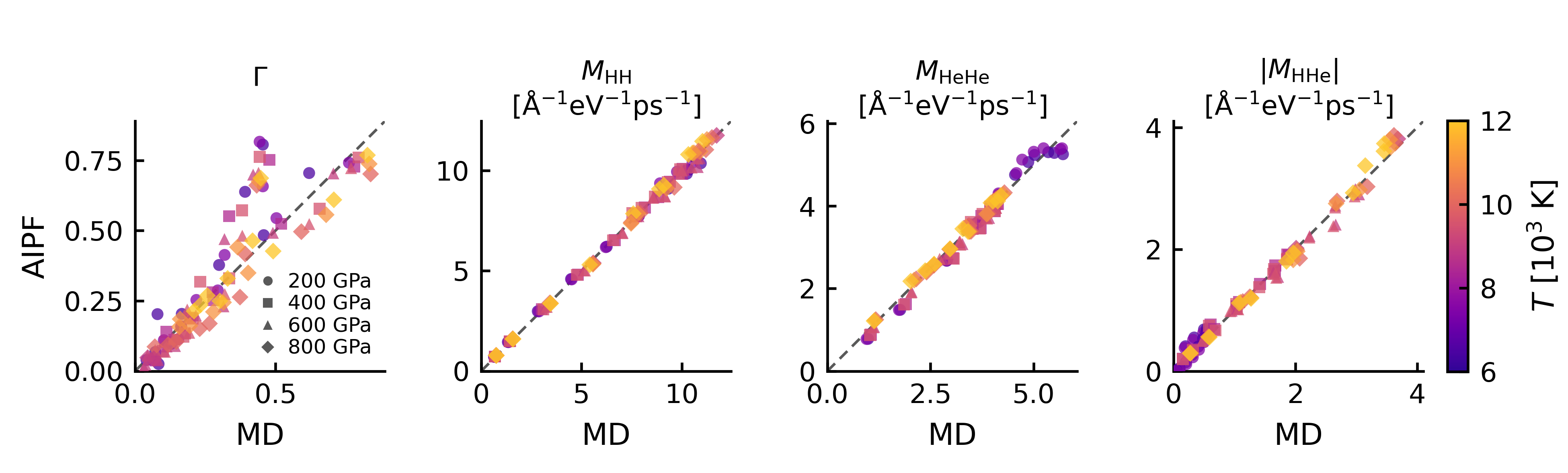}
    \end{minipage}
    \appcaption{\textbf{Agreement with the hydrogen--helium training anchors.}
    The predictions of the AIPF model are compared with molecular-dynamics measurements at the homogeneous states used for the static and mobility anchors. These states span $6500$--$12\,000\,$K at $200$, $400$, $600$ and $800\,$GPa. Marker shapes indicate pressure, colors indicate temperature, and dashed lines denote equality. The first panel compares $\Gamma$ calculated from the learned free-energy curvature with estimates from MD concentration fluctuations. The remaining panels compare the independent components of the symmetric $2\times2$ Onsager mobility matrix, measured from short-time density changes. The cross-species mobility $M_{\mathrm{HHe}}$ is negative at every state and is shown in magnitude.
    \label{fig:appendix_hhe_anchors}}
\appfigend

\end{document}


\title{Supplemental Material for \\ ``Learning \textit{ab initio} phase-field models''}

\author{Mengyi Chen}
\affiliation{Department of Mathematics, National University of Singapore, Singapore}

\author{Peichen Zhong}
\affiliation{Department of Materials Science and Engineering, National University of Singapore, Singapore}
\affiliation{Institute for Functional Intelligent Materials, National University of Singapore, Singapore}

\author{Zihan Zhang}
\affiliation{Department of Materials Science and Engineering, National University of Singapore, Singapore}

\author{Qianxiao Li}
\email{qianxiao@nus.edu.sg}
\affiliation{Department of Mathematics, National University of Singapore, Singapore}
\affiliation{Institute for Functional Intelligent Materials, National University of Singapore, Singapore}

\date{\today}

\maketitle

\tableofcontents

\newpage

\section{Derivation of the mesoscopic model}
\label{sec:si_theory}

\subsection{Mori--Zwanzig formalism}
\label{sec:mori-zwanzig}
We consider a general state space and the autonomous deterministic dynamics on it, 
\begin{equation*}
  \dot{\mathbf Z}(t)=\mathbf G(\mathbf Z(t)),\qquad
  \mathbf Z(0)=\mathbf z,
\end{equation*}
where $\mathbf z$ is a microstate of the state space. 
We are interested in the dynamics of the observable
$\mathbf A(\mathbf z)=(A_1(\mathbf z),\ldots,A_m(\mathbf z))$, and write
$\mathbf A(t)=\mathbf A(\mathbf Z(t))$. The components $A_1,\ldots,A_m$
are assumed functionally independent, with no identity relating them as
functions of $\mathbf z$~\cite{grabert1982projection}. By the chain rule,
\begin{equation*}
  \frac{\mathrm d}{\mathrm dt}\mathbf A(t)
    =\mathbf G(\mathbf Z(t))\cdot\nabla_{\mathbf z}\mathbf A(\mathbf Z(t))
    =(\mathcal L\mathbf A)(\mathbf Z(t)),
  \qquad
  \mathcal Lf(\mathbf z)=\mathbf G(\mathbf z)\cdot\nabla_{\mathbf z}f(\mathbf z),
\end{equation*}
where $\mathcal L$ is the Liouville operator, the generator of the flow on
observables, so that $\mathbf A(t)=(e^{t\mathcal L}\mathbf A)(\mathbf z)$.
Let $\req(\mathbf z)$ be an invariant probability distribution, so that
$\nabla_{\mathbf z}\cdot(\req\mathbf G)=0$ and, for every admissible observable
$f$,
\begin{equation}
  \int\req(\mathbf z)\,\mathcal Lf(\mathbf z)\dd\mathbf z=0.
  \label{eq:invariance}
\end{equation}
\Cref{eq:invariance} follows from the stationarity of $\req$ by the
divergence theorem provided the flux $\req f\mathbf G$ vanishes on the boundary
of the state space and at infinity. For the dynamics used below this holds
because the positions are periodic and $\req$ is compactly supported or decays
at least exponentially in every other coordinate, so every observable of polynomial growth is admissible.
For a fixed value $\mathbf a\in\mathbb R^m$, the Zwanzig projection
maps an observable, also called a phase function, $f(\mathbf z)$ to its
average over $\req$ conditional on
$\mathbf A(\mathbf z)=\mathbf a$,
\begin{equation}
  \begin{aligned}
  \mathcal P_{\mathbf a}f
    &=\frac{1}{\Omega(\mathbf a)}
      \int\req(\mathbf z)\,\delta(\mathbf A(\mathbf z)-\mathbf a)
      f(\mathbf z)\dd\mathbf z,\\
  \Omega(\mathbf a)
    &=\int\req(\mathbf z)\,\delta(\mathbf A(\mathbf z)-\mathbf a)\dd\mathbf z.
  \end{aligned}
  \label{eq:conditional-projection}
\end{equation}
Here $\Omega(\mathbf a)$ is the equilibrium probability density
of the resolved observables. Where $\Omega>0$, the free energy at fixed
temperature $T>0$ is
\begin{equation}
  \mathcal F(\mathbf a)=-\kb T\ln\Omega(\mathbf a).
  \label{eq:free-energy}
\end{equation}
Here $T$ is the temperature of the ensemble $\req$, or any fixed positive
constant when $\req$ is microcanonical. It enters the exact relations below
only through $\mathbf K$ and $\mathcal F$ together, in a combination that does
not depend on it.
For any admissible function $h$ of the resolved observables, conditional
expectation gives
\begin{equation}
  \mathcal P_{\mathbf a}[h(\mathbf A)f]
    =h(\mathbf a)\mathcal P_{\mathbf a}f.
  \label{eq:projection-properties}
\end{equation}
Evaluating the conditional average at $\mathbf a=\mathbf A(\mathbf z)$
defines the projection on observables and its complement:
\begin{equation}
  \mathcal P_{\mathbf A(\mathbf z)}f
    =\left.\mathcal P_{\mathbf a}f\right|_{\mathbf a=\mathbf A(\mathbf z)},
  \qquad
  \mathcal Q_{\mathbf A(\mathbf z)}=1-\mathcal P_{\mathbf A(\mathbf z)},
  \label{eq:projection}
\end{equation}
so that, by \cref{eq:projection-properties},
$\mathcal P_{\mathbf A(\mathbf z)}^2=\mathcal P_{\mathbf A(\mathbf z)}$
and $\mathcal P_{\mathbf A(\mathbf z)}\mathcal Q_{\mathbf A(\mathbf z)}
=\mathcal Q_{\mathbf A(\mathbf z)}\mathcal P_{\mathbf A(\mathbf z)}=0$.
We assume that the operator evolutions $e^{t\mathcal L}$ and
$e^{t\mathcal Q_{\mathbf A(\mathbf z)}\mathcal L}$ exist on the observables used below.

The resolved variables then obey the standard generalized Langevin equation of the projection-operator formalism~\cite{zwanzigMemoryEffectsIrreversible1961,moriTransportCollectiveMotion1965,grabert1982projection,hijonMoriZwanzigFormalism2010a,schillingCoarsegrainedModellingOut2022}.
\begin{theorem}[Generalized Langevin equation (GLE)]
\label{thm:gle}
Under \cref{eq:invariance} and the existence and regularity assumptions
stated above, the resolved trajectory satisfies
\begin{equation}
  \partial_t\mathbf A(t)
  =\mathbf v(\mathbf A(t))
   +\int_0^t\!\dd s\,
    \left[\kb T\,\nabla_{\mathbf a}\cdot\mathbf K(\mathbf a,s)
    -\mathbf K(\mathbf a,s)\nabla_{\mathbf a}\mathcal F(\mathbf a)
    \right]_{\mathbf a=\mathbf A(t-s)}
   +\Rvec(t,\mathbf z),
  \label{eq:free-energy-gle}
\end{equation}
with the drift, the fluctuating force and the memory kernel given by
\begin{equation}
  \mathbf v(\mathbf a)=\mathcal P_{\mathbf a}\mathcal L\mathbf A,
  \qquad
  \Rvec(t,\mathbf z)
  =\big(e^{t\mathcal Q_{\mathbf A(\mathbf z)}\mathcal L}\mathcal Q_{\mathbf A(\mathbf z)}\mathcal L\mathbf A\big)(\mathbf z),
  \label{eq:noise-definition}
\end{equation}
\begin{equation}
  \mathbf K(\mathbf a,s)
  =\frac{1}{\kb T}\,
    \mathcal P_{\mathbf a}
    \left[\Rvec(s,\cdot)\Rvec(0,\cdot)^{\mathsf T}\right].
  \label{eq:kernel-definition}
\end{equation}
The fluctuating force has zero conditional mean
at every admissible resolved value,
\begin{equation}
  \mathcal P_{\mathbf a}\Rvec(t,\cdot)=\mathbf0.
  \label{eq:noise-orthogonality}
\end{equation}
For component indices $\mu,\nu=1,\ldots,m$, the divergence acts on the
second matrix index:
\begin{equation}
  (\nabla_{\mathbf a}\cdot\mathbf K)_\mu
  =\sum_{\nu=1}^m\partial_{a_\nu}K_{\mu\nu}.
  \label{eq:divergence-convention}
\end{equation}
\end{theorem}

\subsection{Application to density fields}
\label{sec:density-fields}

We consider molecular dynamics of $N_\alpha$ atoms of each species $\alpha$ in a periodic cell $D$ of volume $V$, so
$\mathbf z$ comprises the particle positions and momenta. Periodic boundary conditions are assumed throughout this section, so all fields are periodic and integrations by parts over $D$ produce no boundary terms. For Hamiltonian dynamics in the microcanonical (NVE) ensemble this is
the whole state space. For canonical (NVT) dynamics generated by a
Nos\'e--Hoover thermostat~\cite{martynaNoseHooverChains1992},
$\mathbf z$ also contains the thermostat variables, so the state
space is the extended phase space of that scheme, on which the dynamics is
deterministic.

For the density-field specialization below the cell is fixed, so particle
$j$ of species $\alpha$ and mass $m_\alpha$ evolves by
$\dot{\mathbf r}_j^\alpha=\mathbf v_j^\alpha=\mathbf p_j^\alpha/m_\alpha$.
For species $\alpha$ with $N_\alpha$ particles and every reciprocal-lattice
wavevector $\mathbf k$ of the cell $D$, define the retained Fourier coefficients
\begin{equation}
  \hat\rho_{\mathbf k,\alpha}(\mathbf z)
    =\begin{cases}
       \displaystyle\sum_{j=1}^{N_\alpha}e^{-i\mathbf k\cdot\mathbf r_j^\alpha},
       &|\mathbf k|\leq k_{\mathrm{cut}},\\[6pt]
       0,&|\mathbf k|>k_{\mathrm{cut}}.
      \end{cases}
  \label{eq:density-modes}
\end{equation}
Within the cutoff these are the Fourier coefficients of the microscopic number
density $\sum_{j}\delta(\mathbf r-\mathbf r_j^\alpha)$. Since the positions are
real,
\begin{equation}
  \hat\rho_{-\mathbf k,\alpha}=\hat\rho_{\mathbf k,\alpha}^{*},
  \qquad \hat\rho_{\mathbf0,\alpha}=N_\alpha.
  \label{eq:density-mode-constraints}
\end{equation}
Let $\hat{\boldsymbol\rho}(\mathbf z)
=(\hat\rho_{\mathbf k,\alpha}(\mathbf z))_{|\mathbf k|\leq k_{\mathrm{cut}},\,\alpha}$
be the vector of retained coefficients. Along a microscopic trajectory, write
$\hat\rho_{\mathbf k,\alpha}(t)=\hat\rho_{\mathbf k,\alpha}(\mathbf Z(t))$. The
resolved density field is the inverse series of these coefficients, and its
transform returns them,
\begin{equation}
  \begin{aligned}
  \rho_\alpha(\mathbf r,t)
    &=\frac{1}{V}\sum_{\mathbf k}
       \hat\rho_{\mathbf k,\alpha}(t)e^{i\mathbf k\cdot\mathbf r},\\
  \hat\rho_{\mathbf k,\alpha}(t)
    &=\int_D\rho_\alpha(\mathbf r,t)e^{-i\mathbf k\cdot\mathbf r}
       \dd\mathbf r.
  \end{aligned}
  \label{eq:density-fields}
\end{equation}
The sum is finite by \cref{eq:density-modes}, so $\rho_\alpha$ is real and
band-limited, and it shares its retained Fourier coefficients with the
microscopic density while its higher ones vanish. The cutoff sets the spatial
resolution. We write
$\boldsymbol\rho(\mathbf r,t)=(\rho_\alpha(\mathbf r,t))_\alpha$ for the vector
of species density fields. For any field $f$ on $D$, let $\Pi$ keep its
retained Fourier coefficients,
\begin{equation*}
  (\Pi f)(\mathbf r)
    =\frac{1}{V}\sum_{|\mathbf k|\leq k_{\mathrm{cut}}}
      \hat f(\mathbf k)\,e^{i\mathbf k\cdot\mathbf r},
  \qquad
  \hat f(\mathbf k)=\int_D f(\mathbf r)\,e^{-i\mathbf k\cdot\mathbf r}\dd\mathbf r,
\end{equation*}
so that $\boldsymbol\rho=\Pi\boldsymbol\rho$.

We apply \cref{thm:gle} with
$\mathbf A(\mathbf z)=\hat{\boldsymbol\rho}(\mathbf z)$ and write
$\hat{\boldsymbol\rho}(t)=\hat{\boldsymbol\rho}(\mathbf Z(t))$.
The fixed resolved value is still denoted by $\mathbf a$.
The entries of $\hat{\boldsymbol\rho}$ are not functionally independent,
since \cref{eq:density-mode-constraints} ties $\hat\rho_{-\mathbf k,\alpha}$ to
$\hat\rho_{\mathbf k,\alpha}$ and fixes the zero modes. Following
Zubarev and Morozov~\cite{zubarevStatisticalMechanicsNonlinear1983}, the independent variables
are the real and imaginary parts of $\hat\rho_{\mathbf k,\alpha}$ over a
half-space of wavevectors with $\mathbf k\neq\mathbf 0$.
As in Zubarev and Morozov~\cite{zubarevStatisticalMechanicsNonlinear1983}, we assume these variables to be functionally independent, so that the hypotheses of \cref{thm:gle} apply to them.
Every integral over $\mathbf a$ and
every derivative $\partial_{a_\nu}$ below is taken with respect to them,
while the formulas are written in the complex coefficients over the full band
for brevity.
For $|\mathbf k|\leq k_{\mathrm{cut}}$, the chain rule gives
\begin{equation}
  \mathcal L\hat\rho_{\mathbf k,\alpha}
    =-i\mathbf k\cdot\sum_{j=1}^{N_\alpha}
       \mathbf v_j^\alpha e^{-i\mathbf k\cdot\mathbf r_j^\alpha},
  \label{eq:mode-rate}
\end{equation}
which is odd under reversal of all momenta, whereas $\hat{\boldsymbol\rho}$
and $\req$ are even.
Reversing the momenta in \cref{eq:conditional-projection} therefore gives
$\mathcal P_{\mathbf a}\mathcal L\hat{\boldsymbol\rho}
=-\mathcal P_{\mathbf a}\mathcal L\hat{\boldsymbol\rho}$, so the drift
vanishes,
\begin{equation}
  \mathbf v(\mathbf a)
    =\mathcal P_{\mathbf a}\mathcal L\hat{\boldsymbol\rho}
    =\mathbf0.
  \label{eq:drift-vanishes}
\end{equation}
Substituting into \cref{eq:free-energy-gle} yields
\begin{equation}
  \partial_t\hat{\boldsymbol\rho}(t)
    =\int_0^t\!\dd s\,\Big[
      \kb T\,\nabla_{\hat{\boldsymbol\rho}}\cdot
        \mathbf K\big(\hat{\boldsymbol\rho}(t-s),s\big)
      -\mathbf K\big(\hat{\boldsymbol\rho}(t-s),s\big)
        \nabla_{\hat{\boldsymbol\rho}}\mathcal F\big(\hat{\boldsymbol\rho}(t-s)\big)
      \Big]+\Rvec(t,\mathbf z).
  \label{eq:density-gle}
\end{equation}
Here $\mathbf K\big(\hat{\boldsymbol\rho}(t-s),s\big)$ is the memory kernel,
which describes how past density configurations affect the current
evolution of the field. It is a matrix on the independent real density-mode
indices. Derivatives with respect to $\hat{\boldsymbol\rho}$ act on the
resolved-state argument before evaluation at the displayed state, and the
divergence acts on the second matrix index as in
\cref{eq:divergence-convention}.

The fluctuating force $\Rvec(t,\mathbf z)$ and the memory kernel account
for the eliminated momenta and density modes above $k_{\mathrm{cut}}$.
At a fixed initial resolved density $\mathbf a$, the fluctuating
force has zero conditional mean by \cref{eq:noise-orthogonality}, and
satisfies the conditional fluctuation--dissipation
relation~\cite{hijonMoriZwanzigFormalism2010a} of \cref{eq:kernel-definition},
\begin{equation}
  \mathcal P_{\mathbf a}
    \left[\Rvec(s,\cdot)\Rvec(0,\cdot)^{\mathsf T}\right]
  =\kb T\,\mathbf K(\mathbf a,s).
  \label{eq:density-fdt}
\end{equation}
The free energy is defined by \cref{eq:free-energy}
and is a functional of the retained density fields.
These relations are exact under the stated conditions and impose neither
a Markov assumption nor Gaussian noise.

\subsection{From the GLE to the mesoscopic model}
\label{sec:mesoscopic-model}

\Cref{eq:density-gle} is exact, but evaluating its memory kernel and
fluctuating force requires propagating the orthogonal dynamics, which is
generally impractical~\cite{hijonMoriZwanzigFormalism2010a}. We make three assumptions to obtain
a tractable mesoscopic model. We retain the microscopic dynamics of
\cref{sec:density-fields} and the density field defined in
\cref{eq:density-fields}.
First, in many systems, such as simple fluids, fluid mixtures, and colloidal suspensions, the memory kernel decays on a time much shorter than the evolution of the resolved density modes~\cite{ortizdezarateHydrodynamicFluctuations2006,vrugtClassicalDynamicalDensity2020}.
We therefore take the dynamics as Markovian, where the memory kernel is replaced by its time integral~\cite{zwanzigMemoryEffectsIrreversible1961,grabert1982projection,hijonMoriZwanzigFormalism2010a}.

\noindent\textbf{Assumption 1 (Markovian dynamics).}
We assume that the memory kernel and its first derivatives with respect
to the resolved state decay on a time $\tau_{\mathrm{mem}}$ much shorter
than the evolution time $\tau_{\mathrm{field}}$ of every retained density
mode. The time integral below is assumed to converge and to allow
differentiation under the integral with respect to the resolved state.
We define the integrated mobility and approximate the kernel by
\begin{equation}
  \bar{\mathbf M}(\mathbf a)
    =\int_0^\infty\mathbf K(\mathbf a,s)\dd s,
  \qquad
  \mathbf K(\mathbf a,s)
    \approx 2\bar{\mathbf M}(\mathbf a)\delta(s).
  \label{eq:integrated-mobility}
\end{equation}
The delta approximation is understood under time integration against
slowly varying functions. We use the symmetric endpoint convention
$\int_0^\infty\delta(s)\dd s=1/2$, so the one-sided kernel integral
is $\bar{\mathbf M}$.
Only time lags $s\lesssim\tau_{\mathrm{mem}}$ contribute appreciably to
the memory integral. During this interval, the density changes negligibly,
so $\hat{\boldsymbol\rho}(t-s)\approx\hat{\boldsymbol\rho}(t)$.
For $t\gg\tau_{\mathrm{mem}}$, the free-energy contribution therefore becomes
\begin{equation*}
  \begin{aligned}
  &\int_0^t\!\dd s\,
    \mathbf K\big(\hat{\boldsymbol\rho}(t-s),s\big)
    \nabla_{\hat{\boldsymbol\rho}}\mathcal F\big(\hat{\boldsymbol\rho}(t-s)\big)\\
  &\qquad\approx
    \left[\int_0^t\!\dd s\,
      \mathbf K\big(\hat{\boldsymbol\rho}(t),s\big)\right]
    \nabla_{\hat{\boldsymbol\rho}}\mathcal F\big(\hat{\boldsymbol\rho}(t)\big)\\
  &\qquad\approx
    \bar{\mathbf M}\big(\hat{\boldsymbol\rho}(t)\big)
    \nabla_{\hat{\boldsymbol\rho}}\mathcal F\big(\hat{\boldsymbol\rho}(t)\big).
  \end{aligned}
\end{equation*}
The first approximation freezes the density, and the second neglects
the small remaining tail. The same steps apply to the divergence term,
using differentiation under the integral. Thus
\begin{equation*}
  \partial_t\hat{\boldsymbol\rho}(t)
    \approx\kb T\,\nabla_{\hat{\boldsymbol\rho}}\cdot
      \bar{\mathbf M}\big(\hat{\boldsymbol\rho}(t)\big)
    -\bar{\mathbf M}\big(\hat{\boldsymbol\rho}(t)\big)
      \nabla_{\hat{\boldsymbol\rho}}\mathcal F\big(\hat{\boldsymbol\rho}(t)\big)
    +\Rvec(t,\mathbf z).
\end{equation*}
The fluctuation--dissipation relation in \cref{eq:density-fdt}, together
with the delta approximation in \cref{eq:integrated-mobility}, gives
\begin{equation*}
  \mathcal P_{\mathbf a}
    \left[\Rvec(s,\cdot)\Rvec(0,\cdot)^{\mathsf T}\right]
    =\kb T\mathbf K(\mathbf a,s)
    \approx2\kb T\bar{\mathbf M}(\mathbf a)\delta(s).
\end{equation*}

\medskip
Before imposing the local-mobility approximation, we express the memory kernel in terms of correlations of orthogonally propagated currents. By \cref{eq:noise-definition}, $\tilde{\mathbf R}(t)=e^{t\mathcal Q\mathcal L}\mathcal Q\mathcal L\hat{\boldsymbol\rho}$, and $\mathcal Q\mathcal L\hat{\boldsymbol\rho}=\mathcal L\hat{\boldsymbol\rho}-\mathcal P_{\mathbf A(\mathbf z)}\mathcal L\hat{\boldsymbol\rho}=\mathcal L\hat{\boldsymbol\rho}$, since the drift vanishes by \cref{eq:drift-vanishes}. Together with \cref{eq:mode-rate}, written as $\mathcal L\hat\rho_{\mathbf k,\alpha}=-i\mathbf k\cdot\hat{\mathbf j}_{\mathbf k,\alpha}$ where $\hat{\mathbf j}_{\mathbf k,\alpha}=\sum_j\mathbf v_j^\alpha e^{-i\mathbf k\cdot\mathbf r_j^\alpha}$ is the microscopic current, and with the linearity of $e^{t\mathcal Q\mathcal L}$, which commutes with the constant vector $\mathbf k$, this gives $\tilde R_{\mathbf k,\alpha}(t)=-i\mathbf k\cdot\hat{\mathbf J}_{\mathbf k,\alpha}(t)$ with $\hat{\mathbf J}_{\mathbf k,\alpha}(t)=e^{t\mathcal Q\mathcal L}\hat{\mathbf j}_{\mathbf k,\alpha}$ the orthogonally propagated current. Substituting into \cref{eq:kernel-definition} gives the exact form $K_{\mathbf k\alpha,\mathbf k'\beta}(\mathbf a,s)=-\,\mathbf k\cdot\hat{\mathbf C}_{\alpha\beta}(\mathbf k,\mathbf k',s)\cdot\mathbf k'$ with $\hat{\mathbf C}_{\alpha\beta}=(\kb T)^{-1}\mathcal P_{\mathbf a}\big[\hat{\mathbf J}_{\mathbf k,\alpha}(s)\,\hat{\mathbf J}_{\mathbf k',\beta}(0)^{\mathsf T}\big]$,
for every $s\ge0$. Provided these current correlations are time integrable, the integrated mobility of \cref{eq:integrated-mobility} admits the representation
\begin{equation}
  \begin{aligned}
  \bar M_{\mathbf k\alpha,\mathbf k'\beta}
    &=-\,\mathbf k\cdot\hat{\boldsymbol\Lambda}_{\alpha\beta}(\mathbf k,\mathbf k')\cdot\mathbf k',\\
  \hat{\boldsymbol\Lambda}_{\alpha\beta}(\mathbf k,\mathbf k')
    &=\int_0^\infty\hat{\mathbf C}_{\alpha\beta}(\mathbf k,\mathbf k',s)\dd s
     =\frac{1}{\kb T}\int_0^\infty
      \mathcal P_{\mathbf a}\big[\hat{\mathbf J}_{\mathbf k,\alpha}(s)\,\hat{\mathbf J}_{\mathbf k',\beta}(0)^{\mathsf T}\big]\dd s,
  \end{aligned}
  \label{eq:general-mobility}
\end{equation}
where $\hat{\boldsymbol\Lambda}_{\alpha\beta}$ is a $3\times3$ Cartesian tensor that depends on the resolved state $\mathbf a$, and $\mathbf 1$ below denotes the identity on the Cartesian indices. It is the double Fourier transform of a two-point tensor kernel $\boldsymbol\Lambda_{\alpha\beta}(\mathbf r,\mathbf r';[\boldsymbol\rho],T)$, a functional of the whole density field. Explicitly, with $\mathbf J_\alpha(\mathbf r,s)=e^{s\mathcal Q\mathcal L}\sum_j\mathbf v_j^\alpha\delta(\mathbf r-\mathbf r_j^\alpha)$ the orthogonally propagated current field and $\hat{\mathbf J}_{\mathbf k,\alpha}(s)=\int_D\mathbf J_\alpha(\mathbf r,s)e^{-i\mathbf k\cdot\mathbf r}\dd\mathbf r$ its Fourier transform,
\begin{equation}
  \begin{aligned}
  \hat{\boldsymbol\Lambda}_{\alpha\beta}(\mathbf k,\mathbf k')
    &=\iint_D\boldsymbol\Lambda_{\alpha\beta}(\mathbf r,\mathbf r';[\boldsymbol\rho],T)\,
      e^{-i\mathbf k\cdot\mathbf r-i\mathbf k'\cdot\mathbf r'}\dd\mathbf r\dd\mathbf r',\\
  \boldsymbol\Lambda_{\alpha\beta}(\mathbf r,\mathbf r';[\boldsymbol\rho],T)
    &=\frac{1}{\kb T}\int_0^\infty
      \mathcal P_{\mathbf a}\big[\mathbf J_\alpha(\mathbf r,s)\,\mathbf J_\beta(\mathbf r',0)^{\mathsf T}\big]\dd s,
  \end{aligned}
  \label{eq:general-mobility-real}
\end{equation}
where $\boldsymbol\Lambda_{\alpha\beta}$ is the time-integrated correlation of the current field, and the mobility operator acts as $\boldsymbol\mu\mapsto-\nabla\cdot\!\int_D\boldsymbol\Lambda_{\alpha\beta}(\mathbf r,\mathbf r';[\boldsymbol\rho],T)\cdot\nabla'\mu_\beta(\mathbf r')\dd\mathbf r'$. The factors $\mathbf k$ and $\mathbf k'$ express mass conservation, since every term of \cref{eq:density-gle} is a divergence and $\hat\rho_{\mathbf 0,\alpha}=N_\alpha$ is constant. The next assumption restricts only $\hat{\boldsymbol\Lambda}$.

\noindent\textbf{Assumption 2 (Local mobility).}
We restrict $\hat{\boldsymbol\Lambda}$ in \cref{eq:general-mobility} to the local and isotropic form $\boldsymbol\Lambda_{\alpha\beta}(\mathbf r,\mathbf r';[\boldsymbol\rho],T)=M_{\alpha\beta}(\boldsymbol\rho(\mathbf r),T)\,\mathbf 1\,\delta(\mathbf r-\mathbf r')$, that is, $\hat{\boldsymbol\Lambda}_{\alpha\beta}(\mathbf k,\mathbf k')=\hat M_{\alpha\beta}(\mathbf k+\mathbf k')\,\mathbf 1$, so that in the complex coefficients of \cref{sec:density-fields}
\begin{equation}
  \begin{aligned}
  \bar M_{\mathbf k\alpha,\mathbf k'\beta}
    &=-(\mathbf k\cdot\mathbf k')\,
      \hat M_{\alpha\beta}(\mathbf k+\mathbf k'),\\
  \hat{\mathbf M}(\mathbf q)
    &=\int_D\mathbf M(\boldsymbol\rho(\mathbf r),T)
      e^{-i\mathbf q\cdot\mathbf r}\dd\mathbf r.
  \end{aligned}
  \label{eq:local-mobility-class}
\end{equation}
$\bar{\mathbf M}$ is defined only for $|\mathbf k|,|\mathbf k'|\leq k_{\mathrm{cut}}$.
Here $\mathbf M(\boldsymbol\rho,T)$ is the local real-space mobility,
a matrix on species indices.
It is symmetric positive semidefinite and continuously differentiable in
the local densities, with no dependence on density gradients or the field
elsewhere.
This local restriction makes the
integrated mobility easier to identify from trajectory data.
The density field is the one defined
in \cref{eq:density-fields}, and the same retained band is used for every
species.
No locality assumption is imposed on $\mathcal F$.
All mode sums below run over the retained wavevectors, including
both $\mathbf k$ and $-\mathbf k$.

\begin{lemma}[Vanishing thermal drift]
\label{lem:thermal-drift}
Under Assumption~2, the divergence of the mode-space mobility vanishes:
\begin{equation*}
  \nabla_{\hat{\boldsymbol\rho}}\cdot\bar{\mathbf M}=\mathbf0.
\end{equation*}
\end{lemma}
\begin{proof}
\Cref{eq:density-fields} gives
$\partial\rho_\gamma/\partial\hat\rho_{\mathbf k',\beta}
=V^{-1}\delta_{\gamma\beta}e^{i\mathbf k'\cdot\mathbf r}$.
Consequently,
\begin{equation*}
  \frac{\partial\hat M_{\alpha\beta}(\mathbf k+\mathbf k')}
       {\partial\hat\rho_{\mathbf k',\beta}}
    =\frac{1}{V}\int_D
      \frac{\partial M_{\alpha\beta}}{\partial\rho_\beta}
      e^{-i\mathbf k\cdot\mathbf r}\dd\mathbf r
    =\frac{1}{V}
      \widehat{\frac{\partial M_{\alpha\beta}}{\partial\rho_\beta}}(\mathbf k).
\end{equation*}
This expression is independent of $\mathbf k'$, so
\begin{equation*}
  \sum_{\beta,\mathbf k'}
    \frac{\partial\bar M_{\mathbf k\alpha,\mathbf k'\beta}}
         {\partial\hat\rho_{\mathbf k',\beta}}
    =-\frac{1}{V}\sum_\beta
      \widehat{\frac{\partial M_{\alpha\beta}}{\partial\rho_\beta}}(\mathbf k)
      \,\mathbf k\cdot\sum_{\mathbf k'}\mathbf k'
    =\mathbf 0.
\end{equation*}
The last sum vanishes because the retained band contains conjugate
wavevector pairs. The coordinate transformation is linear, so the
divergence also vanishes in the independent real coordinates.
\end{proof}

\medskip
\noindent\textbf{Assumption 3 (Gaussian white noise).}
We assume that the eliminated variables equilibrate rapidly conditional
on the current density, and replace the exact force by a white noise $\hat{\mathbf R}(t)$ in the mode coordinates of $\hat{\boldsymbol\rho}$ that is zero-mean and Gaussian conditional on the current state $\mathbf a$, with conditional covariance
\begin{equation}
  \mathbb E\!\left[\hat{\mathbf R}(t)
           \hat{\mathbf R}(t')^{\mathsf T}\,\middle|\,\mathbf a\right]
    =2\kb T\bar{\mathbf M}(\mathbf a)\delta(t-t'),
  \label{eq:markov-noise-fdt}
\end{equation}

Such a process can be obtained in real space as follows. Let $\boldsymbol\xi$ be a
zero-mean Gaussian space-time white noise with
\begin{equation}
  \mathbb E[\xi_{\alpha i}(\mathbf r,t)\xi_{\beta j}(\mathbf r',t')]
    =\delta_{\alpha\beta}\delta_{ij}
      \delta(\mathbf r-\mathbf r')\delta(t-t'),
  \label{eq:flux-white-noise}
\end{equation}
where $i,j$ are Cartesian components, and set
\begin{equation}
  \mathbf R(\mathbf r,t)
    =\nabla\cdot\left[
       \sqrt{2\kb T\mathbf M(\boldsymbol\rho(\mathbf r),T)}\,
       \boldsymbol\xi(\mathbf r,t)\right],
  \label{eq:real-space-noise}
\end{equation}
with $\boldsymbol\rho$ the field of the current resolved state and the square
root the symmetric positive semidefinite one, which exists by Assumption~2.
Its Fourier coefficients are retained only for $|\mathbf k|\leq k_{\mathrm{cut}}$,
\begin{equation*}
  \hat R_{\mathbf k,\alpha}(t)
    =i\mathbf k\cdot\sum_\gamma\int_D
      e^{-i\mathbf k\cdot\mathbf r}
      \left[\sqrt{2\kb T\mathbf M(\boldsymbol\rho(\mathbf r),T)}
        \right]_{\alpha\gamma}
      \boldsymbol\xi_\gamma(\mathbf r,t)\dd\mathbf r .
\end{equation*}
Conditional on $\mathbf a$, they are zero-mean and jointly Gaussian, being linear in $\boldsymbol\xi$ at fixed $\boldsymbol\rho$, and \cref{eq:flux-white-noise,eq:local-mobility-class} give
\begin{equation*}
  \begin{aligned}
  &\mathbb E\left[\hat R_{\mathbf k,\alpha}(t)
     \hat R_{\mathbf k',\beta}(t')\,\middle|\,\mathbf a\right]\\
  &\qquad=-2\kb T(\mathbf k\cdot\mathbf k')
      \int_D M_{\alpha\beta}(\boldsymbol\rho(\mathbf r),T)
      e^{-i(\mathbf k+\mathbf k')\cdot\mathbf r}\dd\mathbf r\,
      \delta(t-t')\\
  &\qquad=2\kb T\bar M_{\mathbf k\alpha,\mathbf k'\beta}
       (\mathbf a)\delta(t-t'),
  \end{aligned}
\end{equation*}
which is \cref{eq:markov-noise-fdt}.

\medskip
Using \cref{lem:thermal-drift}, the three assumptions give
\begin{equation}
  \partial_t\hat{\boldsymbol\rho}
    =-\bar{\mathbf M}\nabla_{\hat{\boldsymbol\rho}}\mathcal F
      +\hat{\mathbf R}(t),
  \label{eq:mode-ito-model}
\end{equation}
interpreted in the It\^o sense.

\begin{theorem}[Conservative mesoscopic model]
\label{thm:mesoscopic-model}
Under Assumptions 1--3, in real space \cref{eq:mode-ito-model} reads
\begin{equation}
  \partial_t\boldsymbol\rho
    =\Pi\left\{\nabla\cdot\left[
       \mathbf M(\boldsymbol\rho,T)\nabla
       \frac{\delta\mathcal F}{\delta\boldsymbol\rho}\right]
     +\nabla\cdot\left[
       \sqrt{2\kb T\mathbf M(\boldsymbol\rho,T)}\,\boldsymbol\xi\right]\right\}.
  \label{eq:mesoscopic-spde}
\end{equation}
\end{theorem}

\begin{proof}
We compare the retained Fourier coefficients of
\cref{eq:mesoscopic-spde} with the mode equation
\cref{eq:mode-ito-model}. 

\medskip
\begin{samepage}
\noindent\textbf{1. Free-energy derivative.}
Write $\mu_\alpha=\delta\mathcal F/\delta\rho_\alpha$ for the
zero-mean variational derivative in the retained field space. Substituting the Fourier series of an
admissible density variation gives
\begin{equation*}
  \begin{aligned}
  \delta\mathcal F
    &=\sum_\alpha\int_D\mu_\alpha(\mathbf r)
       \delta\rho_\alpha(\mathbf r)\dd\mathbf r\\
    &=\frac{1}{V}\sum_{\alpha,\mathbf q}
       \delta\hat\rho_{\mathbf q,\alpha}
       \int_D\mu_\alpha(\mathbf r)e^{i\mathbf q\cdot\mathbf r}
         \dd\mathbf r\\
    &=\frac{1}{V}\sum_{\alpha,\mathbf q}
       \hat\mu_\alpha(-\mathbf q)\delta\hat\rho_{\mathbf q,\alpha}.
  \end{aligned}
\end{equation*}
\end{samepage}
The positive exponential gives the coefficient at $-\mathbf q$.
Comparing with the differential of $\mathcal F$, with the complex
coefficients treated as independent, yields
\begin{equation}
  \frac{\partial\mathcal F}{\partial\hat\rho_{\mathbf q,\alpha}}
    =\frac{1}{V}\hat\mu_\alpha(-\mathbf q).
  \label{eq:retained-chemical-potential}
\end{equation}

\medskip
\noindent\textbf{2. Deterministic term.}
Substituting \cref{eq:local-mobility-class,eq:retained-chemical-potential}
into the mode drift gives
\begin{equation}
  \begin{aligned}
  &-\sum_{\beta,\mathbf k'}
      \bar M_{\mathbf k\alpha,\mathbf k'\beta}
      \frac{\partial\mathcal F}{\partial\hat\rho_{\mathbf k',\beta}}\\
  &\qquad=\frac{1}{V}\sum_{\beta,\mathbf k'}
      (\mathbf k\cdot\mathbf k')
      \hat M_{\alpha\beta}(\mathbf k+\mathbf k')
      \hat\mu_\beta(-\mathbf k')\\
  &\qquad=-\frac{1}{V}\sum_{\beta,\mathbf q}
      (\mathbf k\cdot\mathbf q)
      \hat M_{\alpha\beta}(\mathbf k-\mathbf q)
      \hat\mu_\beta(\mathbf q),
  \end{aligned}
  \label{eq:mode-drift-substitution}
\end{equation}
where the last line uses $\mathbf q=-\mathbf k'$.
To identify this expression, Fourier transform the real-space flux.
Since $\nabla\mu_\beta=V^{-1}\sum_{\mathbf q}
 i\mathbf q\hat\mu_\beta(\mathbf q)e^{i\mathbf q\cdot\mathbf r}$,
\begin{equation*}
  \begin{aligned}
  \widehat{M_{\alpha\beta}\nabla\mu_\beta}(\mathbf k)
    &=\int_D M_{\alpha\beta}(\boldsymbol\rho(\mathbf r),T)
       \nabla\mu_\beta(\mathbf r)e^{-i\mathbf k\cdot\mathbf r}
       \dd\mathbf r\\
    &=\frac{1}{V}\sum_{\mathbf q}i\mathbf q\hat\mu_\beta(\mathbf q)
       \int_D M_{\alpha\beta}(\boldsymbol\rho(\mathbf r),T)
       e^{-i(\mathbf k-\mathbf q)\cdot\mathbf r}\dd\mathbf r\\
    &=\frac{1}{V}\sum_{\mathbf q}i\mathbf q
       \hat M_{\alpha\beta}(\mathbf k-\mathbf q)
       \hat\mu_\beta(\mathbf q).
  \end{aligned}
\end{equation*}
Taking the divergence multiplies this coefficient by $i\mathbf k\cdot$.
Thus $i^2=-1$ and the sum over $\beta$ give exactly
\cref{eq:mode-drift-substitution}.
Here $\mu_\beta$ lies in the retained band, but the local function
$M_{\alpha\beta}(\boldsymbol\rho,T)$ is not truncated before multiplication.
Its coefficient at $\mathbf k-\mathbf q$ may lie outside that band.

The noise term of \cref{eq:mesoscopic-spde} is \cref{eq:real-space-noise}
by definition. Inverse transformation reconstructs the rate as
\begin{equation}
  \partial_t\rho_\alpha(\mathbf r,t)
    =\frac{1}{V}\sum_{\mathbf k}e^{i\mathbf k\cdot\mathbf r}
      \left[
       i\mathbf k\cdot\sum_\beta
         \widehat{M_{\alpha\beta}\nabla\mu_\beta}(\mathbf k)
       +\hat R_{\mathbf k,\alpha}(t)\right].
  \label{eq:inverse-mode-model}
\end{equation}
This is \cref{eq:mesoscopic-spde}, the projection $\Pi$ coming from the
retained-band indices of $\bar{\mathbf M}$ and $\hat{\mathbf R}$ in
Assumptions~2 and~3.
\end{proof}

In the main text \cref{eq:mesoscopic-spde} is written without $\Pi$, as is
customary for phase-field equations. The projection is nevertheless always in
force, since the state is stored as the retained Fourier coefficients and the
equation is solved in Fourier space, with the nonlinear terms evaluated in real
space and their retained coefficients taken, which is \cref{eq:inverse-mode-model}.

\begin{theorem}[Mass conservation and invariant equilibrium]
\label{thm:mesoscopic-equilibrium}
The model in \cref{eq:mesoscopic-spde} preserves every $N_\alpha$
and admits the invariant probability density
\begin{equation}
  p_{\mathrm{eq}}(\hat{\boldsymbol\rho})
    =\exp\left[-\frac{\mathcal F(\hat{\boldsymbol\rho})}{\kb T}\right]
    {}=\Omega(\hat{\boldsymbol\rho}),
  \label{eq:mesoscopic-equilibrium}
\end{equation}
the equilibrium marginal of \cref{eq:conditional-projection}, normalized by
\cref{eq:free-energy}.
\end{theorem}

\begin{proof}
The integral of a periodic divergence is zero, so
\begin{equation*}
  \dd\int_D\rho_\alpha(\mathbf r,t)\dd\mathbf r=0.
\end{equation*}
Read in the It\^o sense, \cref{eq:mode-ito-model} has drift
$-\bar{\mathbf M}\nabla_{\hat{\boldsymbol\rho}}\mathcal F$ and diffusion matrix
$\kb T\bar{\mathbf M}$, half the covariance in \cref{eq:markov-noise-fdt}. Its
Fokker--Planck equation for the density $p$ of the real mode coordinates is
\begin{equation*}
  \partial_tp
    =\sum_i\partial_i
       \Big[\big(\bar{\mathbf M}
         \nabla_{\hat{\boldsymbol\rho}}\mathcal F\big)_ip\Big]
     +\kb T\sum_{i,j}\partial_i\partial_j\big(\bar M_{ij}p\big)
    =\nabla_{\hat{\boldsymbol\rho}}\cdot
       \left[\bar{\mathbf M}
         \left(p\nabla_{\hat{\boldsymbol\rho}}\mathcal F
               +\kb T\nabla_{\hat{\boldsymbol\rho}}p\right)\right],
\end{equation*}
where $\sum_j\partial_j\bar M_{ij}=0$ by \cref{lem:thermal-drift}.
Substituting \cref{eq:mesoscopic-equilibrium} gives
\begin{equation*}
  p_{\mathrm{eq}}\nabla_{\hat{\boldsymbol\rho}}\mathcal F
    +\kb T\nabla_{\hat{\boldsymbol\rho}}p_{\mathrm{eq}}=\mathbf0,
\end{equation*}
so the probability current vanishes and $\partial_tp_{\mathrm{eq}}=0$.
\end{proof}

\clearpage
\section{Numerical methods}
\label{sec:si_numerics}

We describe the two-species solver used for hydrogen--helium (H--He) and iron--boron (Fe--B), followed by the scalar implementation used for the Lennard-Jones mixture, and then the equilibrium density under an external potential. All mesoscopic simulations reported here include thermal noise. The same integration scheme can also be used for deterministic dynamics by omitting the stochastic contribution.

\subsection{Dynamics}

\paragraph*{Discretization.}
We use a pseudospectral method on a periodic grid and represent the two density fields by their nonredundant Fourier coefficients $\hat{\boldsymbol\rho}(\mathbf k)$.
For simulations initialized from molecular dynamics, we coarse-grain the measured modes, normalize them by the system volume and map them onto the Fourier grid.
The helium-rain simulation starts from the prescribed state described in Section S3.
At each step, we differentiate the local free energy to obtain the local chemical potentials and evaluate the nonlocal contribution as $\widehat{\mathbf W}_\theta(\mathbf k)\hat{\boldsymbol\rho}(\mathbf k)$.
This matrix--vector product replaces direct real-space convolution, and the transformed kernel can be precomputed for a fixed kernel and grid.
We compute the chemical-potential gradients spectrally, multiply them by the local mobility $\mathbf M_\theta(\boldsymbol\rho,T)$ in real space and evaluate the resulting flux divergence spectrally.

\paragraph*{Time integration.}
We use a first-order implicit--explicit (IMEX) scheme based on the reference linear operator
\begin{equation*}
  \mathbf B(\mathbf k)=|\mathbf k|^2\mathbf M_\theta(\bar{\boldsymbol\rho},T)\mathbf H_\theta(\mathbf k),
\end{equation*}
where $\bar{\boldsymbol\rho}$ is the mean initial density and $\mathbf H_\theta(\mathbf k)$ is the Fourier-space Hessian of the learned free-energy functional $\mathcal F_\theta$, evaluated at this uniform reference state.
Writing $\hat{\mathbf N}$ for the Fourier transform of the full deterministic term $\nabla\cdot[\mathbf M_\theta\nabla\boldsymbol\mu]$, we split it as
\begin{equation*}
  \hat{\mathbf N}(\hat{\boldsymbol\rho})
  =-\mathbf B\hat{\boldsymbol\rho}
   +\big[\hat{\mathbf N}(\hat{\boldsymbol\rho})+\mathbf B\hat{\boldsymbol\rho}\big].
\end{equation*}
The first term is treated implicitly and the bracketed remainder explicitly.
The added and subtracted linear terms cancel, so the reference linearization does not replace the full nonlinear dynamics.
Including the stochastic contribution gives
\begin{equation*}
  \hat{\boldsymbol\rho}^{\,n+1}=\hat{\boldsymbol\rho}^{\,n}
   +\Delta t\,\big[\mathbf I+\Delta t\,\mathbf B(\mathbf k)\big]^{-1}
    \big(\hat{\mathbf N}^{\,n}+\hat{\mathbf g}^{\,n}\big),
\end{equation*}
where $\hat{\mathbf N}^{\,n}$ is evaluated at the current density fields and $\hat{\mathbf g}^{\,n}$ is the stochastic contribution defined below.
Each Fourier mode requires only an independent $2\times2$ linear solve, performed analytically, rather than a nonlinear solve over the full spatial grid.
In the absence of noise and additional state projections, the update preserves the stationary solutions of the deterministic dynamics.

\paragraph*{Stochastic contribution.}
At each step, we draw independent standard normal variables $w_{jd}(\mathbf r)$ for each grid cell, species $j$ and Cartesian direction $d$.
We construct the stochastic flux from a Cholesky factor of the local mobility,
\begin{equation*}
  \zeta_{id}(\mathbf r)=\sqrt{\frac{2k_{\mathrm B}T}{\Delta V\,\Delta t}}\;
   \mathsf L_{ij}\big(\boldsymbol\rho(\mathbf r)\big)\,w_{jd}(\mathbf r),
  \qquad
  \mathsf L\mathsf L^{\mathsf T}=\mathbf M_\theta\big(\boldsymbol\rho(\mathbf r),T\big),
\end{equation*}
where $\Delta V$ is the grid-cell volume and repeated species indices are summed.
This construction preserves the cross-species noise correlations specified by the off-diagonal mobility.
We evaluate the flux divergence spectrally and apply the Gaussian filter $\exp(-|\mathbf k|^2\sigma^2/2)$ used to define the coarse-grained fields.
The resulting Fourier contribution is $\hat{\mathbf g}^{\,n}$, whose zero mode is set to zero to conserve the number of each species.
Evaluating the noise amplitude at the beginning of each step gives a semi-implicit Euler--Maruyama discretization in the It\^o convention.

\paragraph*{Lennard-Jones implementation.}
For the Lennard-Jones mixture, we use a scalar version of the same scheme to evolve the composition field, which is stored in real space.
The reference linear operator is the Cahn--Hilliard gradient-energy stabilizer of the semi-implicit Fourier-spectral method~\cite{chenApplicationsSemiimplicitFourierspectral1998,zhuCoarseningKineticsVariablemobility1999},
\begin{equation*}
  B(\mathbf k)=\kappa_{\mathrm{eff}}\,M_{\max}\,|\mathbf k|^4,
\end{equation*}
where $\kappa_{\mathrm{eff}}=\tfrac12\widehat W_\theta''(0)$ is the gradient-energy coefficient given by the small-$k$ expansion of the learned kernel, or the gradient coefficient $\kappa$ of the baselines, and $M_{\max}$ is the maximum of the local mobility over the domain at the current step. It replaces the learned Hessian $\mathbf H_\theta(\mathbf k)$ of the two-species scheme.
The noise construction, Gaussian filtering and zero-mode constraint follow the procedure above, with $\sigma=1.5\,\sigma_{AA}$.

\subsection{Thermodynamics}

The external potential enters $\mathcal F_\theta$ through the coupling term $\int\mathbf V_{\mathrm{ext}}\cdot\boldsymbol\rho\,\mathrm d\mathbf r$ of Methods. At fixed chemical potentials $\boldsymbol\mu$, the equilibrium density minimizes the grand potential,
\begin{equation*}
  \min_{\boldsymbol\rho}\ \Big\{\mathcal F_\theta([\boldsymbol\rho],T)
    -\boldsymbol\mu\!\cdot\!\!\int\!\boldsymbol\rho\,\mathrm d\mathbf r\Big\} .
\end{equation*}
In the canonical ensemble, the particle numbers $\mathbf N$ are fixed and the equilibrium density minimizes $\mathcal F_\theta$ subject to particle conservation,
\begin{equation*}
  \begin{aligned}
      &\min_{\boldsymbol\rho}\ \mathcal F_\theta([\boldsymbol\rho],T) ,\\
      &\text{subject to}\quad \int\!\boldsymbol\rho\,\mathrm d\mathbf r = \mathbf N .
  \end{aligned}
\end{equation*}
Both variational problems give the stationarity condition
\begin{equation*}
  \frac{\delta\mathcal F_\theta([\boldsymbol\rho],T)}{\delta\boldsymbol\rho(\mathbf r)}
    = \boldsymbol\mu \qquad \text{for all } \mathbf r ,
\end{equation*}
which is the stationarity condition of the main text with the external potential retained. The chemical potentials are prescribed in the grand canonical ensemble and arise as Lagrange multipliers enforcing particle conservation in the canonical ensemble.

\clearpage
\section{Experimental details and further results}
\label{sec:si_experiments}

\subsection{Symmetric binary Lennard-Jones mixture}

\label{sec:si_lj}

\paragraph*{Model form.}
The Lennard-Jones mixture is nearly incompressible, so the local total density $\rho_A+\rho_B$ remains $1.0\,\sigma_{AA}^{-3}$ at every point, and the free energy and mobility depend only on the composition field $x_A=\rho_A/(\rho_A+\rho_B)$. We impose symmetry under exchange of species $A$ and $B$. The free energy comprises the parameter-free ideal part and the learned excess part,
\begin{equation*}
\begin{aligned}
  \mathcal F_{\mathrm{id}}&=\int f_{\mathrm{id}}\,\mathrm d\mathbf r,\qquad f_{\mathrm{id}}=T\big[x_A\ln x_A+(1-x_A)\ln(1-x_A)\big],\\
  \mathcal F_{\mathrm{exc},\theta}&=\int f_{\mathrm{loc},\theta}(x_A,T)\,\mathrm d\mathbf r
  +\tfrac12\iint\big[x_A(\mathbf r)-\tfrac12\big]\,W_\theta(\lvert\mathbf r-\mathbf r'\rvert)\,\big[x_A(\mathbf r')-\tfrac12\big]\,\mathrm d\mathbf r\,\mathrm d\mathbf r',\\
  f_{\mathrm{loc},\theta}&=u_{0,\theta}(x_A)+T\,u_{1,\theta}(x_A).
\end{aligned}
\end{equation*}%
The local energy $u_{0,\theta}$ is an even Taylor polynomial in $x_A-\tfrac12$, with terms of orders $2$, $4$, $6$ and $8$. The function $u_{1,\theta}$ is an input-convex neural network with one input and width $32$, symmetrized under $x_A\to1-x_A$. The temperature-independent radial kernel $W_\theta(r)$ is represented by a multilayer perceptron with two layers of width $16$, multiplied by a quintic envelope that vanishes at $R_{\mathrm{cut}}=3.0\,\sigma_{AA}$.

The mobility is $M_\theta(x_A,T)=\gamma\,x_A(1-x_A)\,c_\theta(T)$, where $c_\theta(T)=\exp[-E_a(1/T-1/T_{\mathrm{ref}})]$ and $T_{\mathrm{ref}}=1.65$. Both $\gamma$ and $E_a$ are learned.

\paragraph*{Training and validation data.}
The datasets and simulation conditions are summarized in \cref{tab:si_lj_runs}. Slab trajectories provide the dynamical training data. The trajectories at $T=1.20$ are held out from training and model selection and used only for validation. Homogeneous cubes provide the mobility and bulk anchors at four temperatures, $T=1.50$, $1.60$, $1.70$ and $1.80$, with measurements averaged over three independent seeds at each temperature. Quench trajectories are excluded from training and used only to assess the predicted coarsening dynamics.

The dynamical residual compares an estimate of the time derivative over a window of frames with the model drift averaged over the same window. For fields $\hat{\boldsymbol\rho}_j$ saved at interval $\Delta$ and a window centered at frame $j_c$, the time derivative is estimated as
\begin{equation}
  \widehat{\partial_t\hat{\boldsymbol\rho}}(t_{j_c})
  =\frac{1}{w\Delta}\Big[\frac1w\sum_{m=1}^{w}\hat{\boldsymbol\rho}_{j_c+m}
   -\frac1w\sum_{m=0}^{w-1}\hat{\boldsymbol\rho}_{j_c-m}\Big].
  \label{eq:si_drift_estimator}
\end{equation}
This estimator divides the difference between the mean fields over the following and preceding $w$ frames by $\delta t=w\Delta$, the interval defined in the main text. We evaluate the model drift at $n_s$ evenly spaced frames within the same window and take a weighted average. The dynamical residual is the difference between this averaged drift and the estimated time derivative.

For the Lennard-Jones mixture, frames are saved every $\tau$, and $w=25$ gives $\delta t=25\,\tau$. Window centers are separated by $20$ frames, and the drift is evaluated at $n_s=5$ frames per window. The loss averages over all windows and training trajectories. To prevent strongly driven trajectories from dominating, trajectory $r$ is weighted by $w_r\propto1/(m_r+\bar m)$, where $m_r$ is the mean squared band-limited $H^{-1}$ norm of its estimated time derivative and $\bar m$ is the median across trajectories.

The training loss is $\mathcal L=\mathcal L_{\mathrm{dyn}}+\lambda_M\mathcal L_M+\lambda_{\mathrm{bulk}}\mathcal L_{\mathrm{bulk}}$, with unit weight for the dynamical term and $\lambda_M=\lambda_{\mathrm{bulk}}=0.01$. We use neither a finite-wavevector static anchor nor an isobar anchor, and impose no additional constraint term for this symmetric mixture at fixed total density.

\begin{table}[!t]
  \centering
  \footnotesize
  \renewcommand{\arraystretch}{1.25}
  \setlength{\tabcolsep}{6pt}
  \begin{tabular}{@{}lccc@{}}
    \toprule
    & Slab & Homogeneous cube & Quench cube \\
    \midrule
    Particles $N$ & $3456$ & $4000$ & $6912$ \\
    Cells & $6\times6\times24$ & $10^3$ & $12^3$ \\
    Temperatures & $1.00$--$1.70$ & $1.50, 1.60, 1.70, 1.80$ & $1.10$--$1.30$ \\
    Production ($\tau$) & $20\,000$ & $5000$ & $10\,000$ \\
    \bottomrule
  \end{tabular}
  \caption{Microscopic simulations used for the Lennard-Jones mixture. All three datasets use the force-shifted pair potential described in Methods, with $\epsilon_{AB}=0.5\,\epsilon_{AA}$, total density $\rho=1.0\,\sigma_{AA}^{-3}$ and mean composition $x_A=0.5$. Trajectories are generated by overdamped Langevin dynamics with $\gamma_t=2.0$ and a time step of $2\times10^{-4}\,\tau$. Configurations are saved every $\tau$. Slab trajectories provide the dynamical training data, with $T=1.20$ held out from both training and model selection. Homogeneous cubes provide the bulk and mobility anchors. Quench cubes are used exclusively to validate spinodal decomposition. Two additional systems of $55\,296$ particles are simulated with an underdamped integrator under checkerboard and hexagonal external potentials. These runs sample equilibrium density profiles and are excluded from training.
  \label{tab:si_lj_runs}}
\end{table}

\paragraph*{Anchor measurements.}
Configurations are saved every $\tau$. We estimate the mobility from the covariance of density-mode increments at a lag of $\tau$, averaged over $0.4\le k\le1.0\,\sigma_{AA}^{-1}$, after analytically deconvolving the Gaussian coarse-graining filter. For the bulk anchor, we fit the low-$k$ concentration structure factor to the Ornstein--Zernike form and extrapolate to obtain $S_{cc}(0)$. At each temperature, the anchor compares the learned free-energy curvature at equimolar composition with $\rho\,k_{\mathrm B}T/S_{cc}(0)$, weighted by the standard error of the intercept across seeds. The bulk anchor therefore constrains only the zero-wavevector curvature at $x_A=0.5$.

\begin{figure}[!t]
  \centering
  \panelw{0.578\linewidth}{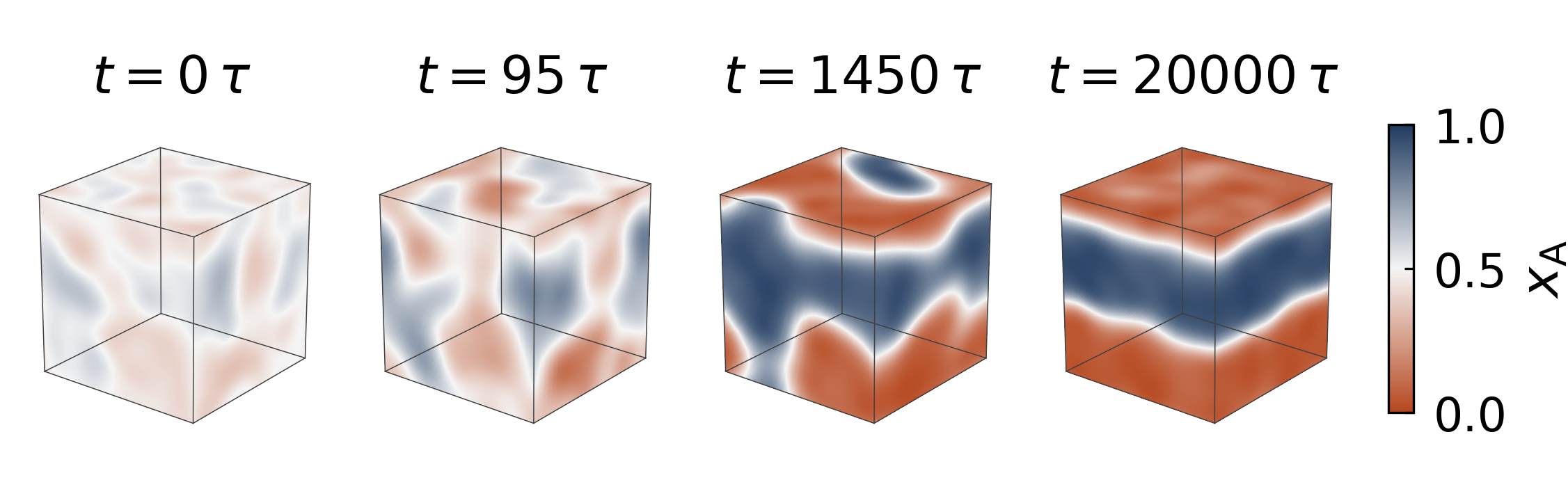}{a}\hfill
  \panelw{0.398\linewidth}{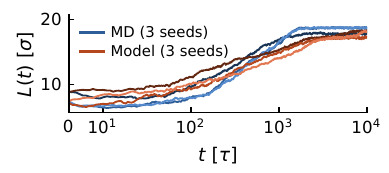}{b}\\[4pt]
  \panelw{0.578\linewidth}{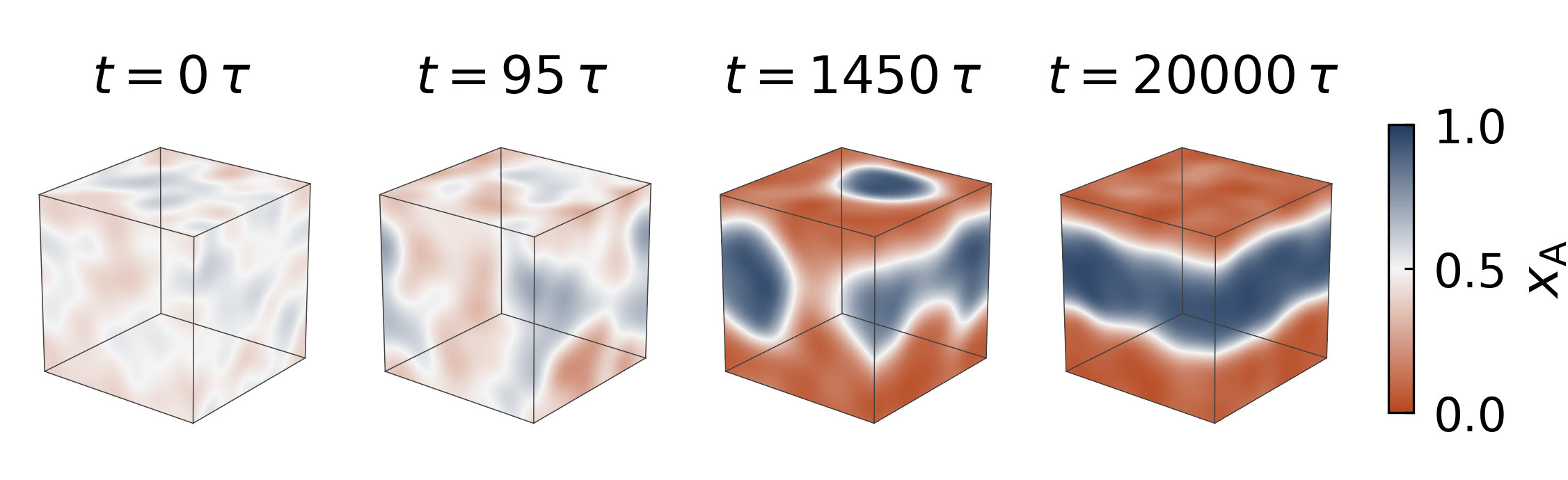}{c}\hfill
  \panelw{0.398\linewidth}{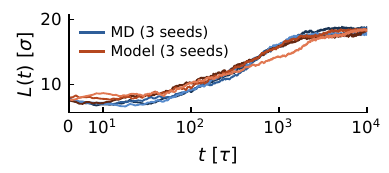}{d}\\[4pt]
  \panelw{0.578\linewidth}{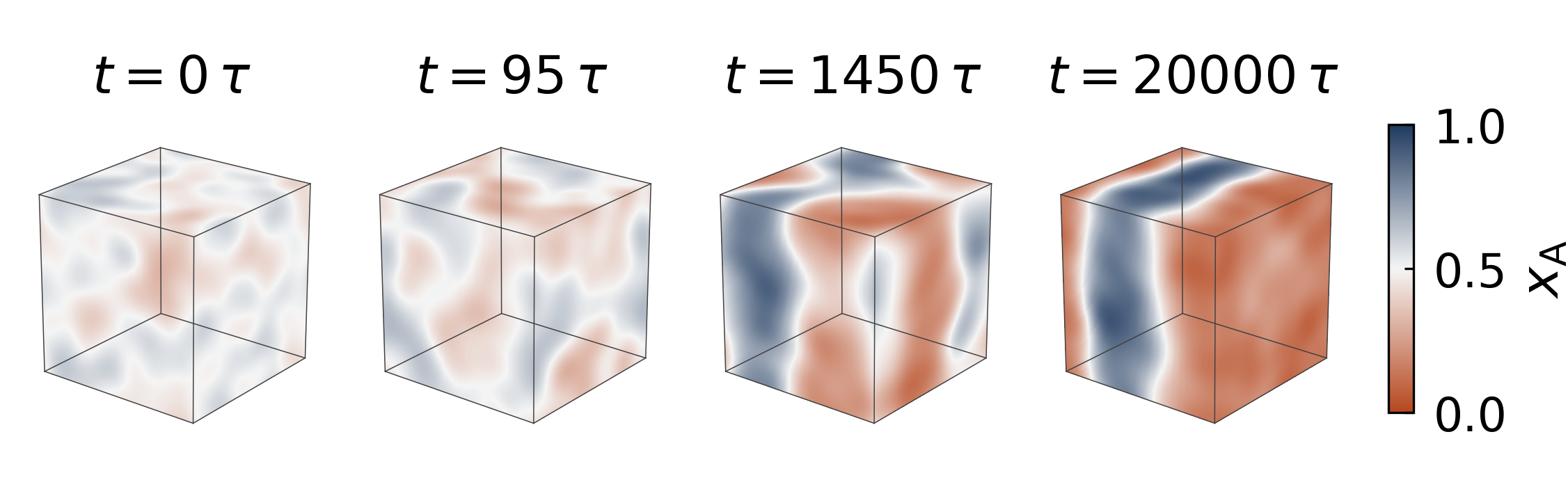}{e}\hfill
  \panelw{0.398\linewidth}{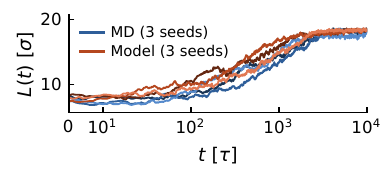}{f}\\[4pt]
  \panelw{0.578\linewidth}{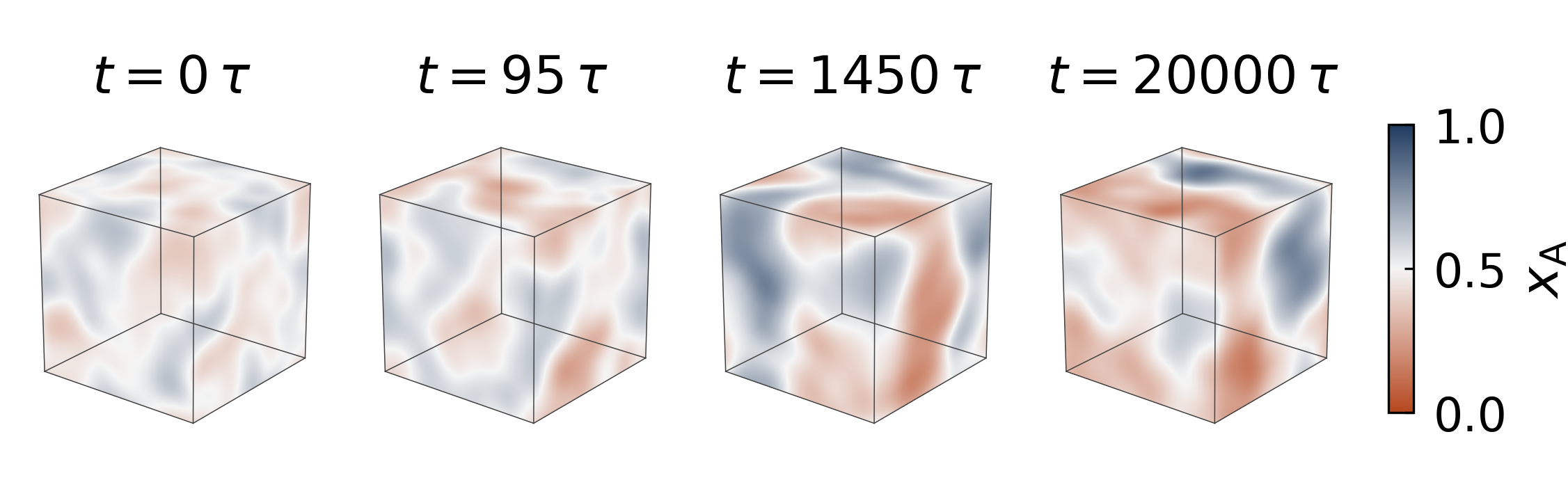}{g}\hfill
  \panelw{0.398\linewidth}{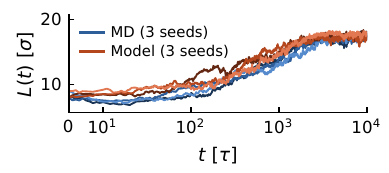}{h}
  \caption{\textbf{Coarsening of the Lennard-Jones mixture.}
  Results at $T=1.10$, $1.15$, $1.25$ and $1.30$ are shown from top to bottom, complementing the results at $T=1.20$ in the main text.
  \textbf{a,c,e,g}, Composition fields from the \textit{ab initio} phase-field (AIPF) model at four times.
  \textbf{b,d,f,h}, Domain size $L(t)$, defined in \cref{eq:si_Lt}, compared between three independent molecular dynamics (MD) runs and three independent AIPF runs.
  \label{fig:si_lj_coarsening}}
\end{figure}

\paragraph*{Domain growth.}
We characterize spinodal decomposition by a length scale obtained from the inverse first moment of the composition spectrum,
\begin{equation}
  L(t)=2\pi\,\frac{\sum_{\mathbf k}S(\mathbf k,t)}
                  {\sum_{\mathbf k}|\mathbf k|\,S(\mathbf k,t)},
  \label{eq:si_Lt}
\end{equation}
where $S(\mathbf k,t)=|\hat\psi(\mathbf k,t)|^2$. For molecular dynamics, $\hat\psi=\tfrac12(\hat\rho_{\mathbf k,A}-\hat\rho_{\mathbf k,B})$, with the Gaussian coarse-graining filter applied analytically to the bare particle modes. For the \textit{ab initio} phase-field (AIPF) model, $\hat\psi$ is the Fourier amplitude of the already coarse-grained composition field. This construction gives the same estimator for both descriptions. The sums include individual modes on the box's wavevector lattice over $0<|\mathbf k|\le1.5\,\sigma_{AA}^{-1}$. We avoid shell averaging because replacing wavevectors by shell centers imposes an artificial upper limit on $L(t)$ that is reached during these runs.

We extract the growth exponent $\beta$ from the slope of $\log L$ against $\log t$, using a window after the initial transient and before finite-size saturation. For each method, we select the half-decade with the steepest slope within this interval. The quoted uncertainty represents the scatter across independent seeds. At the present box size, the local slope does not reach a plateau. The fitted $\beta$ is therefore an effective, window-dependent exponent rather than an asymptotic growth law. Distinguishing candidate asymptotic laws would require a box several times larger.

We evaluate the growth exponent at five quench temperatures between $1.10$ and $1.30$ (\cref{fig:si_lj_coarsening}). Agreement is closest at $T=1.20$, the case shown in the main text. The largest discrepancy occurs at $T=1.10$, where the AIPF model gives $\beta=0.161\pm0.020$, compared with $0.293\pm0.024$ from molecular dynamics.

\FloatBarrier

\subsection{Iron--boron melt}

\paragraph*{Model form.}
We describe the melt by the species-density fields $\boldsymbol\rho=(\rho_{\mathrm B},\rho_{\mathrm{Fe}})$. The free-energy functional comprises the parameter-free ideal contribution $k_{\mathrm B}T\sum_\alpha\int\rho_\alpha(\ln\rho_\alpha-1)\,\mathrm d\mathbf r$, with $\rho_\alpha$ in \AA$^{-3}$, and the learned excess contribution
\begin{equation*}
  \begin{aligned}
  \mathcal F_{\mathrm{exc},\theta}&=\int f_{\mathrm{loc},\theta}(\boldsymbol\rho,T)\,\mathrm d\mathbf r
  +\tfrac12\iint\boldsymbol\rho^{\!\top}(\mathbf r)\,\mathbf W_\theta(\lvert\mathbf r-\mathbf r'\rvert)\,\boldsymbol\rho(\mathbf r')\,\mathrm d\mathbf r\,\mathrm d\mathbf r'.
  \end{aligned}
\end{equation*}
The ideal contribution differs from the form given in Methods by $k_{\mathrm B}T\sum_\alpha\int\rho_\alpha\ln\Lambda_\alpha^3\,\mathrm d\mathbf r$, where $\Lambda_\alpha$ is the thermal wavelength of species $\alpha$. This term is linear in the densities at fixed temperature and changes neither the pressure nor the free energy of mixing.
The local function $f_{\mathrm{loc},\theta}$ is a multilayer perceptron of the two densities and the temperature, with architecture $3\to64\to64\to64\to64\to1$ and GELU activations.
The temperature-independent, symmetric $2\times2$ kernel contains three radial components corresponding to B--B, B--Fe and Fe--Fe interactions. Each component is represented by a $1\to16\to16\to1$ perceptron of the separation, multiplied by a quintic envelope that vanishes at $R_{\mathrm{cut}}=4.5\,$\AA{}.

The mobility is parametrized as $\mathbf M_\theta=\mathbf C_\theta^{1/2}\mathbf D^{1/2}\mathbf L_\theta\mathbf L_\theta^{\!\top}\mathbf D^{1/2}\mathbf C_\theta^{1/2}$, with $\mathbf D=\mathrm{diag}(\rho_{\mathrm B},\rho_{\mathrm{Fe}})$ and $\mathbf C_\theta=\mathrm{diag}\big(c_{\mathrm B,\theta}(T),c_{\mathrm{Fe},\theta}(T)\big)$. A $2\to16\to16\to3$ perceptron of
the densities supplies the lower-triangular factor $\mathbf L_\theta$, which makes $\mathbf M_\theta$ positive semidefinite by construction. The temperature dependence is $c_{\alpha,\theta}(T)=\exp[-E_\alpha(1/k_{\mathrm B}T-1/k_{\mathrm B}T_{\mathrm{ref}})]$, with $T_{\mathrm{ref}}=1800\,$K and one learned activation energy $E_\alpha$ for each species.

\paragraph*{Training data and state selection.}
The molecular-dynamics datasets cover $0$, $5$ and $10\,$GPa, with the same grid of $145$ state points at each pressure (\cref{tab:si_feb_runs}). The temperatures are $T\in\{1200,\allowbreak 1300,\allowbreak 1400,\allowbreak 1500,\allowbreak 1600,\allowbreak 1700,\allowbreak 1800,\allowbreak 1900,\allowbreak 2000,\allowbreak 2300,\allowbreak 2600\}\,$K. At each temperature, the boron fractions are $x_{\mathrm B}\in\{0.10,\allowbreak 0.20,\allowbreak 0.30,\allowbreak 0.40,\allowbreak 0.50,\allowbreak 0.60,\allowbreak 0.65,\allowbreak 0.70,\allowbreak 0.75,\allowbreak 0.80,\allowbreak 0.85,\allowbreak 0.90,\allowbreak 0.95\}$, with additional points at $x_{\mathrm B}=0.98$ for $1500$ and $1800\,$K.

We classify each state from the concentration structure factor $S_{cc}(k,t)$. A state is spinodal when $\Gamma_{\mathrm{MD}}\le0.35$ and coarsening is confirmed on a trajectory of at least $200\,$ps. Coarsening requires that the peak of $S_{cc}(k)$ either moves toward the smallest wavevector of the box or stays at a finite wavevector while its amplitude grows, and that the peak exceeds five times the background without declining at the end of the trajectory. A state is stable when three conditions hold. First, $\Gamma_{\mathrm{MD}}\ge1$. Second, the Ornstein--Zernike estimate of $S_{cc}(0)$ rests on at least $100$ independent samples of the long-wavelength composition modes, which corresponds to a relative error of $10\,\%$. Third, the long-wavelength band $k\le0.52\,$\AA$^{-1}$ shows no growth over the second half of the trajectory, meaning that the late-to-early change in its amplitude is within $1.5$ standard deviations and the series is stationary. All other states are classified as unresolved.

States are excluded from all trajectory-based loss terms if $T\le1300\,$K or if the slower species diffuses less than one bond length over the trajectory, as determined by $6\min(D_{\mathrm B},D_{\mathrm{Fe}})\,t<9\,$\AA$^2$. The remaining trajectories contribute to the dynamical loss. Stable states that also satisfy the anchor-selection threshold provide the static and mobility anchors. Separate equation-of-state calculations provide the isobar anchor.

\begin{table}[!t]
  \centering
  \footnotesize
  \renewcommand{\arraystretch}{1.25}
  \setlength{\tabcolsep}{6pt}
  \begin{tabular}{@{}lccc@{}}
    \toprule
    & $0\,$GPa & $5\,$GPa & $10\,$GPa \\
    \midrule
    State points & $145$ & $145$ & $145$ \\
    Runs & $156$ & $159$ & $149$ \\
    Production (ps) & \multicolumn{3}{c}{$50$ to $500$, per state point} \\
    Spinodal / stable / unresolved & $12$ / $88$ / $45$ & $8$ / $81$ / $56$ & $4$ / $72$ / $69$ \\
    Excluded from training & $31$ & $36$ & $41$ \\
    Equation-of-state rows & $208$ & $106$ & $128$ \\
    \bottomrule
  \end{tabular}
  \caption{Molecular-dynamics datasets for the iron--boron melt. Trajectories are generated for homogeneous cubes of $3456$ atoms in the isotropic isothermal--isobaric (NPT) ensemble using the fine-tuned MACE potential of Zhang \textit{et al.}~\cite{zhangInterfacialmeltStabilityThermodynamic2026}. Each system is melted at $2600\,$K for $10\,$ps before simulation at the target temperature. The time step is $1\,$fs, and configurations are saved every $0.1\,$ps. The state-point grid and selection criteria are described in the text. Retained trajectories provide the dynamical training data, and stable states that satisfy the anchor-selection threshold also provide the static and mobility anchors. Separate equation-of-state calculations use $512$ atoms, with $5\,$ps each for melting, equilibration and averaging, and provide the isobar anchor.
  \label{tab:si_feb_runs}}
\end{table}

\paragraph*{Training objective and constraints.}
The dynamical residual uses
the estimator of \cref{eq:si_drift_estimator},
with $w=100$ frames saved every $0.1\,$ps, giving $\delta t=10\,$ps. Window centers are spaced every $40$ frames, and $n_s=5$. All trajectories carry equal weight. The model is evaluated at states midway between consecutive frames.

The loss is $\mathcal L=\lambda_{\mathrm{dyn}}\sum_s\mathcal L_{\mathrm{dyn}}^{(s)}+\lambda_M\mathcal L_M+\lambda_S\mathcal L_S+\lambda_{\mathrm{bulk}}\mathcal L_{\mathrm{bulk}}+\lambda_P\mathcal L_P+\lambda_W\mathcal L_W$. The dynamical term sums the contributions from nonequilibrium and stable runs at all three pressures, with $\lambda_{\mathrm{dyn}}=1.5\times10^4$. The remaining weights are $\lambda_M=0.5$, $\lambda_S=0.03$, $\lambda_{\mathrm{bulk}}=0.01$, $\lambda_P=5.0$ and $\lambda_W=0.3$. The hinge penalty $\mathcal L_W$ penalizes violations of the lower bound $\kappa k^2$ on $\widehat{\mathbf W}_\theta(k)-\widehat{\mathbf W}_\theta(0)$, with $\kappa=2.0$.

We impose no convexity penalty on the local free energy and no prior on the thermodynamic factor. This allows the local free energy to become nonconvex in the boron-rich region where the instability is observed.

The isobar anchor compares the model's Euler pressure with the target pressure of each equation-of-state state point,
\begin{equation}
  \mathcal L_P=\Big\langle
    \Big(\frac{P_\theta(\boldsymbol\rho,T)-P_{\mathrm{meas}}}{\max(|P_{\mathrm{meas}}|,P_{\mathrm{ref}})}\Big)^{\!2}
  \Big\rangle_{\boldsymbol\rho,T} ,
  \label{eq:si_LP_feb}
\end{equation}
where $P_{\mathrm{ref}}=10\,$GPa and the denominator avoids division by zero at ambient pressure. All three target pressures lie at or below $P_{\mathrm{ref}}$, so the same denominator applies to every state and the penalty measures the squared absolute pressure residual in units of $P_{\mathrm{ref}}$.

\paragraph*{Anchor measurements.}
Configurations are saved every $0.1\,$ps.
We evaluate the apparent mobility at lags of one to ten frames, fit its diffusive branch linearly in $\Delta t$ for each wavevector shell and extrapolate to $\Delta t=0$. The resulting shell intercepts are then extrapolated linearly in $k^2$ to obtain the zero-wavevector mobility. The static anchor uses measured shells up to $k=1.5\,$\AA$^{-1}$, whereas the Ornstein--Zernike fit uses $k<0.8\,$\AA$^{-1}$.

\begin{figure}[!htb]
  \centering
  \includegraphics[width=\linewidth]{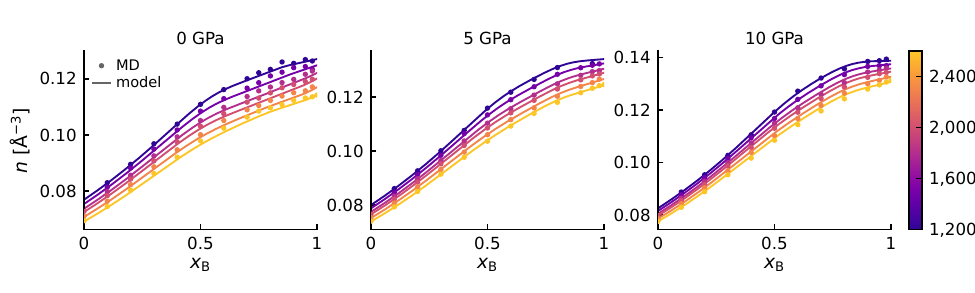}
  \caption{Equation of state of the iron--boron melt at $0$, $5$ and $10\,$GPa. Number density $n$ is shown as a function of boron fraction $x_{\mathrm B}$ along six isotherms between $1200$ and $2600\,$K, distinguished by color. Points denote molecular-dynamics reference states obtained with the fine-tuned potential. Lines show the learned isobars, calculated by solving the model pressure equation at fixed composition and temperature. The $442$ reference states are included in the isobar anchor during training, so this comparison assesses the fit to that constraint.
  \label{fig:si_feb_nx}}
\end{figure}

\paragraph*{Equation of state.}
We assess the fit to the isobar constraint using the $442$ reference states (\cref{fig:si_feb_nx}). At each reference composition and temperature, we solve the model pressure equation by bisection to obtain the density at the target pressure. 

\label{sec:si_feb}

\FloatBarrier

\subsection{Hydrogen--helium mixture}

\paragraph*{Model form.}
We describe the mixture by the species-density fields $\boldsymbol\rho=(\rho_{\mathrm H},\rho_{\mathrm{He}})$. The free-energy functional comprises the parameter-free ideal contribution $k_{\mathrm B}T\sum_\alpha\int\rho_\alpha(\ln\rho_\alpha-1)\,\mathrm d\mathbf r$, with $\rho_\alpha$ in \AA$^{-3}$, and the learned excess contribution
\begin{equation*}
  \begin{aligned}
  \mathcal F_{\mathrm{exc},\theta}&=\int f_{\mathrm{loc},\theta}(\boldsymbol\rho,T)\,\mathrm d\mathbf r
  +\tfrac12\iint\boldsymbol\rho^{\!\top}(\mathbf r)\,\mathbf W_\theta(\lvert\mathbf r-\mathbf r'\rvert)\,\boldsymbol\rho(\mathbf r')\,\mathrm d\mathbf r\,\mathrm d\mathbf r',\\
  f_{\mathrm{loc},\theta}&=u_{0,\theta}(\boldsymbol\rho)+k_{\mathrm B}T\,u_{1,\theta}(\boldsymbol\rho)+k_{\mathrm B}T\ln\frac{T}{T_{\mathrm{ref}}}\,u_{2,\theta}(\boldsymbol\rho).
  \end{aligned}
\end{equation*}
The ideal contribution uses the same convention as for the iron--boron melt. The local functions $u_{0,\theta}$, $u_{1,\theta}$ and $u_{2,\theta}$ are multilayer perceptrons with architecture $2\to h\to h\to1$, GELU activations and widths $h=16$, $32$ and $16$, respectively. The reference temperature is $T_{\mathrm{ref}}=8000\,$K. Before training, the basis function $k_{\mathrm B}T\ln(T/T_{\mathrm{ref}})$ is orthogonalized against $1$ and $k_{\mathrm B}T$ over $9000$--$12\,000\,$K. This redistributes its constant and linear parts among $u_{0,\theta}$, $u_{1,\theta}$ and $u_{2,\theta}$ without changing the space spanned by the basis functions. The temperature-independent, symmetric $2\times2$ kernel $\mathbf W_\theta$ contains three radial components corresponding to H--H, H--He and He--He interactions. Each component is represented by a $1\to16\to16\to1$ perceptron of the separation, multiplied by a quintic envelope that vanishes at $R_{\mathrm{cut}}=4.5\,$\AA{}.

The mobility is parametrized as $\mathbf M_\theta=\mathbf D^{1/2}\mathbf L_\theta\mathbf L_\theta^{\!\top}\mathbf D^{1/2}$, where $\mathbf D=\mathrm{diag}(\rho_{\mathrm H},\rho_{\mathrm{He}})$ and a perceptron of the densities supplies the lower-triangular factor $\mathbf L_\theta$. This parametrization makes $\mathbf M_\theta$ positive semidefinite by construction. No explicit temperature factor is included.

\paragraph*{Training data.}
The AIPF model is trained on molecular-dynamics trajectories generated with a machine-learning potential at $200$, $400$, $600$ and $800\,$GPa. The datasets comprise slabs, homogeneous cubes and equation-of-state calculations, as summarized in \cref{tab:si_hhe_runs}. Slabs and cubes provide the dynamical training data. Cubes that satisfy the temperature and single-phase criteria described below also provide the static, bulk and mobility anchors. All $484$ equation-of-state rows contribute to the isobar anchor.

\begin{table}[!ht]
  \centering
  \footnotesize
  \renewcommand{\arraystretch}{1.25}
  \setlength{\tabcolsep}{5pt}
  \begin{tabular}{@{}lccc@{}}
    \toprule
    & Slabs & Homogeneous cubes & Equation of state \\
    \midrule
    Atoms & $2000$--$4700$ & $3456$ & $512$ \\
    Ensemble & NPT, $z$ only & NPT, isotropic & NPT, isotropic \\
    Time step (fs) & $0.1$ & $0.2$ & $0.2$ \\
    Production (ps) & $40$ & $30$ & $1$ \\
    Runs at $200$, $400$, $600$, $800\,$GPa & $48$, $56$, $64$, $88$ & $79$, $90$, $101$, $133$ & $121$ rows each \\
    \bottomrule
  \end{tabular}
  \caption{Molecular-dynamics datasets for hydrogen--helium. All runs use the MACE potential of Wang, Hamel and Cheng~\cite{wangHydrogenheliumImmiscibilityBoundary2026a}, a Nos\'e--Hoover thermostat and a Martyna--Tobias--Klein barostat. Configurations are saved every $0.02\,$ps. Slabs are initialized with a double hyperbolic-tangent composition profile containing two interfaces in a $12\times12\times40\,$\AA{} box, with pressure control applied only along the interface normal. Cubes are melted for $1.5\,$ps at $2000\,$K above the target temperature before production. At $800\,$GPa, cube simulations span $2000$--$12\,000\,$K in increments of $1000\,$K and helium fractions of $0.05$--$0.98$. At each lower pressure, they span helium fractions of $0.05$--$0.95$ at temperatures on both sides of the critical line. Slabs and cubes contribute to the dynamical loss. The $98$ cube runs that meet the temperature and single-phase criteria also provide the static, bulk and mobility anchors. The $484$ equation-of-state rows provide the isobar anchor and the density paths for the thermodynamic-factor prior.
  \label{tab:si_hhe_runs}}
\end{table}

At each pressure and temperature, we simulate slabs with eight composition pairs. The temperature grids start at $2000\,$K with increments of $1000\,$K and contain $6$, $7$, $8$ and $11$ temperatures at $200$, $400$, $600$ and $800\,$GPa, respectively. This gives $48$, $56$, $64$ and $88$ slab runs. At the three lower pressures, the composition pairs are $(0.05,0.50)$, $(0.05,0.95)$, $(0.15,0.85)$, $(0.25,0.75)$, $(0.35,0.65)$, $(0.40,0.60)$, $(0.45,0.55)$ and $(0.50,0.95)$. At $800\,$GPa, where the miscibility gap is widest, we use pairs closer to the pure components, $(0.00,0.90)$, $(0.00,1.00)$, $(0.10,0.80)$, $(0.10,0.90)$, $(0.10,1.00)$, $(0.20,0.80)$, $(0.20,0.90)$ and $(0.30,0.70)$. Slab trajectories contribute only to the dynamical loss.

The dynamical residual uses
the estimator of \cref{eq:si_drift_estimator},
with $w=25$ frames saved every $0.02\,$ps, giving $\delta t=0.5\,$ps. Window centers are spaced every $10$ frames, and $n_s=5$. Trajectories are weighted as for the Lennard-Jones mixture.

\begin{figure}[!ht]
  \centering
  \includegraphics[width=\linewidth]{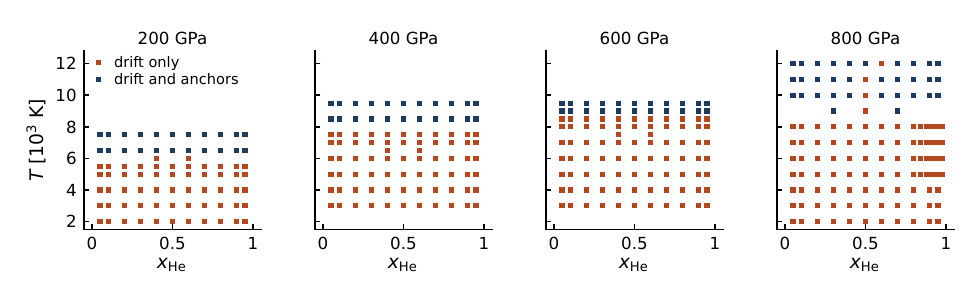}
  \caption{Coverage of the hydrogen--helium cube datasets at the four training pressures. Each run is shown at its helium fraction and temperature. Rust denotes runs used only for the dynamical loss. Navy denotes runs that also provide the static, bulk and mobility anchors. Anchor selection requires temperatures above $6100$, $7800$, $8800$ and $8973\,$K at $200$, $400$, $600$ and $800\,$GPa, respectively, and single-phase behavior over the sampled interval. Runs below these thresholds are excluded from the anchors because the uncertainty in the concentration structure factor is underestimated. The datasets contain $79$, $90$, $101$ and $133$ cube runs at the four pressures, of which $22$, $22$, $22$ and $32$ provide anchor data.
  \label{fig:si_hhe_coverage}}
\end{figure}

The cube datasets contain $79$, $90$, $101$ and $133$ runs at the four pressures. Their roles in training are shown in \cref{fig:si_hhe_coverage}. Anchor data are restricted to temperatures above $6100$, $7800$, $8800$ and $8973\,$K, respectively, because uncertainties in the concentration structure factor are underestimated below these thresholds. Runs must also remain single phase over the sampled interval. These criteria select $22$, $22$, $22$ and $32$ cube runs for the anchors, giving $98$ in total. All other cube runs contribute only to the dynamical loss.

\paragraph*{Training objective and anchor measurements.}
The loss is $\mathcal L=\sum_{s=1}^{8}\mathcal L_{\mathrm{dyn}}^{(s)}+\lambda_M\mathcal L_M+\lambda_S\mathcal L_S+\lambda_{\mathrm{bulk}}\mathcal L_{\mathrm{bulk}}+\lambda_P\mathcal L_P+\lambda_{\mathrm{conv}}\mathcal L_{\mathrm{conv}}+\lambda_\Gamma\mathcal L_\Gamma$. The eight dynamical terms correspond to the slab and cube datasets at each of the four pressures and enter with unit weight. The remaining weights are $\lambda_M=0.5$, $\lambda_S=0.01$, $\lambda_{\mathrm{bulk}}=0.01$, $\lambda_P=5.0$, $\lambda_{\mathrm{conv}}=1.0$ and $\lambda_\Gamma=1.0$.
The weights were set by a coarse sweep over a few values of each, judged by the fit to the anchors, the equation of state and the dynamical residual on the training data.

Configurations are saved every $0.02\,$ps. We evaluate the apparent mobility at lags of one to ten frames and exclude lags shorter than $0.06\,$ps, where fast nonhydrodynamic fluctuations introduce a lag-independent contribution to the increment variance. For each wavevector shell, we fit the diffusive branch of the apparent mobility linearly in $\Delta t$ and extrapolate to $\Delta t=0$. We then obtain $\mathbf M(0)$ by fitting the shell intercepts to a universal kernel shape over $k\le1.6\,$\AA$^{-1}$, with one fitted parameter. The static anchor uses measured shells up to $k=1.5\,$\AA$^{-1}$, whereas $S_{cc}(0)$ is obtained from an Ornstein--Zernike fit over $k\le0.9\,$\AA$^{-1}$.

The isobar anchor compares the model's Euler pressure with the target pressure of each equation-of-state state point,
\begin{equation}
  \mathcal L_P=\Big\langle
    \Big(\frac{P_\theta(\boldsymbol\rho,T)-P_{\mathrm{meas}}}{P_{\mathrm{meas}}}\Big)^{\!2}
  \Big\rangle_{\boldsymbol\rho,T} .
  \label{eq:si_LP}
\end{equation}
Normalizing each residual by its target pressure gives a relative-error penalty, so that the weight $\lambda_P$ has the same interpretation at all four pressures.

\paragraph*{Functional constraints.}
Two additional one-sided penalties constrain the shape of the learned functional. They encode modeling priors and vanish wherever their respective conditions are satisfied.

The first penalizes deviations from convexity of the local free energy. Let $\lambda_{\min}$ be the smallest eigenvalue of $\partial^2 f_{\mathrm{loc},\theta}/\partial\boldsymbol\rho\,\partial\boldsymbol\rho$ and $m$ a fixed margin. We define
\begin{equation*}
  \mathcal L_{\mathrm{conv}}=\big\langle\,\big[\max(0,\,m-\lambda_{\min})\big]^{2}\,\big\rangle .
\end{equation*}
The penalty is zero for $\lambda_{\min}\ge m$ and grows quadratically below this threshold. We estimate the average by Monte Carlo sampling. At each optimization step, we draw new points uniformly within the density trust region and uniformly in logarithmic temperature over the training range. The nonlocal kernel is excluded from this penalty. Enforcing local convexity assigns the bulk demixing instability to $\widehat{\mathbf W}_\theta(0)$ rather than to the local free energy. This separation is a modeling choice.

The second penalty constrains the composition dependence of the thermodynamic factor,
\begin{equation*}
  \mathcal L_\Gamma=\big\langle\,\big[\min(0,\,\partial^2\Gamma/\partial x^2)\big]^{2}\,\big\rangle .
\end{equation*}
It penalizes negative curvature of $\Gamma(x)$, with no additional margin. We evaluate $\Gamma$ from the local Hessian and $\widehat{\mathbf W}_\theta(0)$ along measured equation-of-state density paths. We compute its second derivative using a five-point stencil. The average is restricted to the interior of the sampled composition range. Dilute compositions are excluded to avoid penalizing their physical curvature. Pressure and temperature paths are resampled at each optimization step.

\paragraph*{Helium-rain simulation.}
The helium-rain panels in the main text are obtained from a single stochastic simulation of the learned model using the Fourier pseudospectral scheme described in Section S2. We use a $192\times192\times512$ grid with spacing $0.70\,$\AA{} and periodic boundaries in all three directions. The full computational domain measures $13.4\times13.4\times35.8\,$nm and corresponds to approximately $4.4$ million atoms. The gravitational potential and prescribed temperature profile are extended by reflection along the vertical direction and smoothed near the reflection planes to maintain periodicity. No reflection symmetry is imposed on the density fields or stochastic noise, so these planes do not impose reflecting boundary conditions. The main-text panels show one half of the periodic domain, measuring $13.4\times13.4\times17.9\,$nm and corresponding to approximately $2.2$ million atoms.

The initial density fields are uniform, with protosolar composition $x_{\mathrm{He}}=0.089$ and total number density $0.6753\,$\AA$^{-3}$, corresponding to the equilibrium density at $400\,$GPa. Within the displayed half-domain, the fixed temperature profile increases from $3700\,$K at the top to $3900\,$K at the bottom and is linear away from the $8\,$\AA{} smoothing layers at each end. The entire domain is initially metastable, with both ends inside the binodal and outside the spinodal. The temperature profile thus produces a gradient in supersaturation without crossing a phase boundary. Gravity is imposed through an external potential linear in height, with $g=2.68\times10^{-3}\,$eV\,amu$^{-1}$\,\AA$^{-1}$, approximately $10^{14}$ times Saturn's gravity. This value gives a Bond number of $4.00$ over the free height of $163.2\,$\AA{}, which is the physical height of $179.2\,$\AA{} minus an $8\,$\AA{} layer at each end over which gravity is smoothly switched off, matching the ratio of gravitational to interfacial forces in a planetary rain zone thousands of kilometers deep.

We integrate the model using the stochastic scheme described in Section S2, with a time step of $5\times10^{-3}\,$ps and the noise amplitude set by the fluctuation--dissipation relation. Density fields are saved every $2\,$ps. We use a single noise seed and show the first nanosecond in the main-text panels.

\FloatBarrier

\clearpage
\bibliographystyle{apsrev4-2}
\bibliography{references_arxiv}